\documentclass[11pt]{article}
\usepackage{fullpage}
\usepackage{cite}
\usepackage{graphicx}
\usepackage{amsmath}
\usepackage{amssymb}
\title{}
\date{}

\def\beq{\begin{equation}}
\def\eeq{\end{equation}}
\allowdisplaybreaks
\begin{document}
\bibliographystyle{utphys}
\newcommand{\cN}{{\cal N}} 
\newcommand{\msbar}{\ensuremath{\overline{\text{MS}}}}
\newcommand{\DIS}{\ensuremath{\text{DIS}}}
\newcommand{\abar}{\ensuremath{\bar{\alpha}_S}}
\newcommand{\bb}{\ensuremath{\bar{\beta}_0}}
\newcommand{\rc}{\ensuremath{r_{\text{cut}}}}
\newcommand{\Nd}{\ensuremath{N_{\text{d.o.f.}}}}
\setlength{\parindent}{0pt}

\def\theequation{\thesection.\arabic{equation}}

\titlepage
\begin{flushright}
BRX-TH-667\\
BOW-PH-157\\
\end{flushright}

\vspace*{0.5cm}

\begin{center}
{\Large \bf Wilson line approach to gravity in the high energy limit}

\vspace*{1cm}
\textsc{S. Melville$^a$\footnote{s.melville.1@research.gla.ac.uk}, S. G. Naculich$^b$\footnote{naculich@bowdoin.edu}, H. J. Schnitzer$^c$\footnote{schnitzr@brandeis.edu} and C. D. White$^a$\footnote{Christopher.White@glasgow.ac.uk} } \\

\vspace*{0.5cm} $^a$ SUPA, School of Physics and Astronomy, University of Glasgow,\\ Glasgow G12 8QQ, Scotland, UK\\

\vspace*{0.5cm} $^b$ Department of Physics, Bowdoin College, Brunswick, ME 04011, USA\\

\vspace*{0.5cm} $^c$ Theoretical Physics Group, Martin Fisher School of Physics, \\ Brandeis University, Waltham, MA 02454, USA\\

\end{center}

\vspace*{0.5cm}

\begin{abstract}
We examine the high energy (Regge) limit of gravitational scattering using a 
Wilson line approach previously used in the context of non-Abelian gauge 
theories. Our aim is to clarify the nature of the Reggeization of the graviton 
and the interplay between this Reggeization and the so-called eikonal phase 
which determines the spectrum of gravitational bound states. Furthermore, we 
discuss finite corrections to this picture. Our results are of relevance to 
various supergravity theories, and also help to clarify the relationship 
between gauge and gravity theories. 
\end{abstract}

\vspace*{0.5cm}

\section{Introduction}
\setcounter{equation}{0}

The structure of scattering amplitudes in both gauge and gravity theories 
continues to attract significant attention, due to a wide variety of 
phenomenological and formal applications. Although superficially very 
different from each other, there is mounting evidence that gauge and gravity
theories may be related to each other in intriguing ways. 
Such developments motivate the need to study aspects
of amplitudes in a variety of theories using a common language, and to compare 
and contrast phenomena in gauge theories with their gravitational 
counterparts.\\

\begin{figure}
\begin{center}
\scalebox{0.8}{\includegraphics{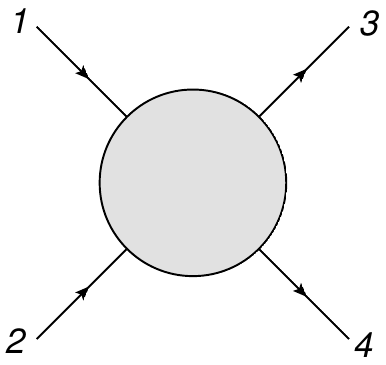}}
\caption{Particle labels used throughout for $2\rightarrow2$ scattering.}
\label{fig:2to2}
\end{center}
\end{figure}

This paper studies one such phenomenon, that of four-point scattering of 
massive (and massless) particles in the so-called {\it Regge limit}, in which 
the center-of-mass energy 
far exceeds the momentum transfer. The properties of amplitudes
in this limit have been studied for a long time, for example in the context
of strong interactions predating the advent of QCD (see 
e.g. refs.~\cite{Collins:1977jy,Landshoff,Gribov} and references therein). Their
asymptotic high energy behavior is dictated by singularities in the complex
angular momentum plane, which may take the form of poles or cuts. Simple poles
give rise to a power-like growth of scattering amplitudes with the 
center-of-mass energy:
\begin{equation}
{\cal A}\sim \left(\frac{s}{-t}\right)^{\alpha(t)},
\label{amppower}
\end{equation}
where we have defined the Mandelstam invariants\footnote{In this paper,
we use the metric convention $(+,-,-,-)$.} 
\begin{equation}
s=(p_1+p_2)^2;\quad t=(p_1-p_3)^2;\quad u=(p_1-p_4)^2.
\label{mandies}
\end{equation}
These satisfy the momentum conservation constraint
\begin{equation}
s+t+u=\sum_{i=1}^4 m_i^2
\label{momcon}
\end{equation}
in terms of the particle momenta $\{p_i\}$ and masses $\{m_i\}$, and we label
particles as shown in figure~\ref{fig:2to2}. The function $\alpha(t)$ in 
eq.~(\ref{amppower}) is known as the {\it Regge trajectory}, whose physical 
origin is the exchange of a family of particles in the $t$-channel. 
Reggeization has also been studied within the context of perturbative quantum
field theory, for both scalar and (non)-Abelian gauge 
theories~\cite{Mandelstam:1965zz,Abers:1967zz,
McCoy:1976ff,Frolov:1970ij,
Gribov:1970ik,Cheng:1969bf,Balitsky:1979ap,Bogdan:2006af,Tyburski:1975mr,
Lipatov:1976zz,Mason:1976fr,Cheng:1977gt,Fadin:1975cb,
Kuraev:1977fs,Kuraev:1976ge,Mason:1976ky,
Sen:1982xv,Fadin:1977jr,Fadin:1995xg,
Fadin:1996tb,Fadin:1995km,Blumlein:1998ib,DelDuca:2001gu,Bogdan:2002sr,
Fadin:2006bj}. There, Regge behavior of amplitudes follows after first 
demonstrating that elementary constituents themselves {\it Reggeize}. For,
example, one may show in QCD that the Feynman gauge propagator in the Regge
limit is dressed according to
\begin{equation}
-\frac{\eta_{\mu\nu}}{k^2}\rightarrow -\frac{\eta_{\mu\nu}}{k^2}\left(\frac{s}{-t}\right)^{\tilde{\alpha}(t)},
\label{gluonreg}
\end{equation}
where $\tilde{\alpha}(t)$ is related in a straightforward way to the Regge 
trajectory $\alpha(t)$. The gluon and quark trajectories in QCD are known to 
two-loop order~\cite{Fadin:1995xg,Fadin:1996tb,Fadin:1995km,Blumlein:1998ib,
DelDuca:2001gu,Bogdan:2002sr}. At one loop order, they are given by
\begin{equation}
  \tilde{\alpha}^{(1)}(t) \, = \frac{\alpha_s(\mu^2)}{2\pi}\, \left(\frac{\mu^2}{-t}\right)^\epsilon \frac{C_R}{\epsilon}, 
\label{alpha1}
\end{equation}
in $d= 4-2\epsilon$ dimensions, where $C_R$ is the quadratic Casimir operator in 
the appropriate representation, and $\mu$ the renormalization scale. Note that 
this is purely infrared singular (up to scale-related logarithms). Apart from 
the particle-dependent Casimir, there is a universal coefficient, which may be
written in terms of the one-loop cusp anomalous 
dimension~\cite{Korchemskaya:1994qp,Korchemskaya:1996je}. The latter quantity
controls the ultraviolet renormalization of Wilson line operators, and this
connection will become clear in what follows. At two-loop order, the Regge
trajectories of the quark and gluon are no longer purely infrared singular,
also involving finite terms. \\

\begin{figure}
\begin{center}
\scalebox{0.8}{\includegraphics{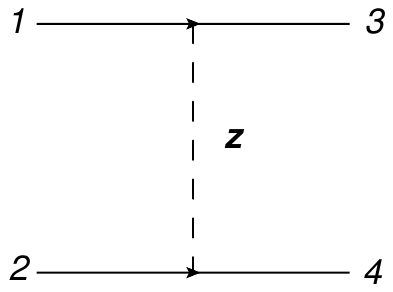}}
\caption{The Regge limit as two Wilson lines separated by a transverse distance $\vec{z}$.}
\label{fig:Wilson}
\end{center}
\end{figure}

A convenient formalism for studying Reggeization was introduced 
in refs.~\cite{Korchemskaya:1994qp,Korchemskaya:1996je}, and is based upon the fact 
that in the Regge limit of $2\rightarrow2$ scattering, the incoming particles
glance off each other, such that the outgoing particles are highly forward,
and essentially do not recoil. They can therefore only change by a phase,
and for this phase to have the right gauge-transformation properties to form
part of a scattering amplitude, it must correspond to a Wilson line operator.
Thus, scattering in the forward limit can be described as two Wilson lines 
separated by a transverse distance (or {\it impact parameter}) $\vec{z}$, a 
situation depicted in figure~\ref{fig:Wilson}. The ultraviolet behavior of 
the Wilson line correlator reproduces the infrared singularities of the 
scattering amplitude, and thus the infrared singular parts of the gluon Regge 
trajectory. Ultraviolet renormalization of Wilson lines is governed by the 
cusp anomalous dimension, hence the connection between this quantity and the 
Regge trajectory noted above. \\

The relationship between infrared singularities and the Regge limit
was studied further recently in
refs.~\cite{Bret:2011xm,DelDuca:2011ae}, which used a conjectured
formula for the all-order infrared (IR) singularity structure of QCD
(the {\it dipole formula} of
refs.~\cite{Gardi:2009qi,Becher:2009cu,Becher:2009qa}, itself
motivated by explicit two-loop
calculations~\cite{Sterman:2002qn,Aybat:2006wq, Aybat:2006mz}) to show
that Reggeization occurs generically up to next-to-leading logarithmic
order in $s/(-t)$, for any allowable $t$-channel exchange. The
corresponding Regge trajectory is dictated by the cusp anomalous
dimension, as already noted in
refs.~\cite{Korchemskaya:1994qp,Korchemskaya:1996je}, and involves the
quadratic Casimir operator in the representation of the exchanged
particle. Beyond this logarithmic order, the authors of
refs.~\cite{Bret:2011xm,DelDuca:2011ae} noted a breakdown of simple
Regge pole behavior, associated with a color operator which has also
been linked to the breakdown of collinear factorization in certain
circumstances~\cite{Catani:2011st,Catani:2012iw,Forshaw:2012bi}.  A
breakdown of simple Regge pole behavior at this order is consistent
with previous two-loop calculations of quark and gluon
scattering~\cite{DelDuca:2008pj}, and is likely to signal the
appearance of Regge cuts associated with multi-Reggeon
exchange~\cite{Kuraev:1976ge,Bartels:1999xt}. The analysis of
refs.~\cite{Bret:2011xm,DelDuca:2011ae} also used the Regge limit to
constrain possible corrections to the QCD dipole formula (known to
break already for massive
particles~\cite{Becher:2009kw,Ferroglia:2009ep,
  Ferroglia:2009ii}). Such corrections may potentially occur at
three-loop order, and have also been investigated in
refs.~\cite{Dixon:2009ur,
  Becher:2009qa,Vernazza:2011aa,Ahrens:2012qz,Naculich:2013xa} (see
also refs.~\cite{Laenen:2008gt,Mitov:2010rp,Chien:2011wz,Gardi:2010rn,
  Gardi:2011wa,Gardi:2011yz,Gardi:2013ita,Dukes:2013wa} for recent
work on understanding IR singularities at higher orders).\\

Although much is known about the Regge limit of gauge theories in perturbative
gauge theory, the situation in gravity is more confused\footnote{Throughout 
this paper, we use the term {\it gauge theory} to refer only to Abelian
and non-Abelian gauge symmetries acting on internal degrees of freedom, 
rather than on spacetime degrees of freedom as in gravity.}. The one-loop Regge
trajectory of the graviton was first derived in refs.~\cite{Lipatov:1982it,Lipatov:1982bf,Lipatov:1982vv}, within the context of both Einstein-Hilbert gravity, and
its supersymmetric extensions.  
One of the authors of the present paper (HJS)
argued in ref.~\cite{Schnitzer:2007kh} 
(based on earlier studies employing analyticity arguments~\cite{Grisaru:1973vw,
Grisaru:1973ku,Grisaru:1974cf,Grisaru:1981ra,Grisaru:1982bi}) 
that Reggeization of the graviton in ${\cal N}=8$ supergravity follows from that of 
the gluon in ${\cal N}=4$ super-Yang-Mills 
theory~\cite{Drummond:2007aua,Naculich:2007ub}
(see also refs.~\cite{DelDuca:2008pj,Naculich:2009cv}) 
as a consequence of the well-known KLT relations~\cite{Kawai:1985xq} relating 
scattering amplitudes in the two theories. This was considered from a Feynman 
diagrammatic point of view in ref.~\cite{Schnitzer:2007rn,Henn:2010bk,Henn:2010ir}, 
which also discussed 
the potential structure of Regge cut contributions in supergravity. \\

There has recently been a rekindled interest in the Regge limit of gravity. A
chief motivation is the study of gravitational scattering in the transplanckian
regime~\cite{Giddings:2010pp} (see also ref.~\cite{'tHooft:1987rb,
Verlinde:1991iu,Amati:1987uf,Amati:1990xe,Amati:1992zb,Amati:1993tb}). This 
regime allows one to explore conceptual questions of quantum gravity, such as 
the existence, or otherwise, of a gravitational 
S-matrix~\cite{Giddings:2009gj,Giddings:2011xs}. An interesting feature is 
that high energy scattering in gravity is dominated by long-distance rather 
than short-distance behavior, a fact which is ultimately traceable to the
dimensionality of the gravitational coupling constant, and the masslessness
of the graviton. The lack, or otherwise, 
of ultraviolet renormalizability ceases to be problematic in this limit, and
one may show that the long-distance behavior is insensitive to the amount of
supersymmetry. However, some confusion remained in 
ref.~\cite{Giddings:2010pp} about the role of graviton Reggeization, and 
the interplay between this and the so-called {\it eikonal phase} which appears 
at high energy~\cite{Kabat:1992tb}, and which is associated with the formation 
of gravitational bound states. This confusion is in part related to the fact
that the graviton Regge trajectory is linear in the squared momentum 
transfer~$t$~\cite{Lipatov:1982it,Lipatov:1982bf,Lipatov:1982vv,
Schnitzer:2007kh,Schnitzer:2007rn}, and thus becomes kinematically subleading 
in the strict Regge limit $s/(-t)\rightarrow\infty$. The issue of graviton 
Reggeization is also complicated by double logarithmic contributions in 
$s/(-t)$, which have been discussed at length in ref.~\cite{Bartels:2012ra}.\\

The Regge limit has also been the focus of studies which aim to relate the
properties of gauge and gravity theories. Examples 
include refs.~\cite{Saotome:2012vy,Vera:2012ds}, which use the high energy limit to
probe the all-order validity of the proposed double copy structure between
gauge and gravity theories~\cite{Bern:2008qj,Bern:2010ue,Bern:2010yg}. As the
work of refs.~\cite{Korchemskaya:1994qp,Korchemskaya:1996je} (and, 
subsequently, refs.~\cite{Bret:2011xm,DelDuca:2011ae}) makes clear, the Regge limit
can be at least partially understood in terms of soft gluon physics and Wilson
lines. The soft limit of gravity was first considered 
in ref.~\cite{Weinberg:1965nx}, and has recently been more extensively studied 
in refs.~\cite{Naculich:2011ry,White:2011yy,Akhoury:2011kq,Beneke:2012xa,
Miller:2012an,Oxburgh:2012zr,BoucherVeronneau:2011nm}. The latter papers seek 
to cast the gravitational behavior in terms of contemporary gauge theory 
language, and thus to expose common physics in the soft limits of both 
theories. This includes the introduction of Wilson line operators for soft 
graviton emission~\cite{Naculich:2011ry,White:2011yy},
whose vacuum expectation values give rise to a 
gravitational {\it soft function}, the UV singularities of which correspond to 
the IR singularities of a scattering amplitude. As in QCD, this function 
exponentiates. Unlike QCD, however, the gravitational soft function has the 
special property of being {\it one-loop exact}, meaning that there are 
no higher loop corrections to the exponent~\cite{Naculich:2011ry,White:2011yy,
Akhoury:2011kq}.\\

The aim of this paper is to examine the Regge limit of (super-)gravity using
Wilson lines, using a similar approach to the QED / QCD case 
of refs.~\cite{Korchemskaya:1994qp,Korchemskaya:1996je}. There are a number of 
motivations for doing so. 
First, the analysis presents an interesting 
application of the gravitational Wilson line operators 
of refs.~\cite{Naculich:2011ry,White:2011yy}. 
Second, the calculation provides a 
common language for Reggeization in both gauge and gravity theories, which is
particularly elegant in revealing common features of the two cases (such as the
appearance of relevant quadratic Casimir operators in Regge trajectories). 
Third, the Wilson line calculation ties together a number of previous results
in gravity in a particularly transparent fashion, and helps to clarify some
of the confusions inherent in the existing literature (such as the interplay
between the eikonal phase and Reggeization of the graviton). We will
also discuss the impact of infrared-finite corrections to the scattering 
amplitude, using one- and two-loop results in a variety of supergravity 
theories~\cite{Green:1982sw,Dunbar:1994bn,Dunbar:1995ed,Bern:1998ug,Naculich:2008ew,Brandhuber:2008tf,Bern:2011rj,BoucherVeronneau:2011qv}. \\

The structure of the paper is as follows. In section~\ref{sec:QCD}, we review
the approach of refs.~\cite{Korchemskaya:1994qp,Korchemskaya:1996je} to the 
Regge limit in terms of Wilson lines, with some slight differences to which we
draw attention.
In section~\ref{sec:gravity} we carry out
a similar calculation in quantum gravity, using the Wilson line operators 
of refs.~\cite{Naculich:2011ry,White:2011yy}, and compare the results with the QCD
case. In section~\ref{sec:finite}, we examine the impact of finite terms
in various supergravity theories on the interpretation of the scattering 
amplitude in the Regge limit. In section~\ref{sec:multi}, we apply the Wilson
line approach to multigraviton scattering, for any number of gravitons.  
Finally, in section~\ref{sec:conclude} we discuss
our results before concluding. Certain technical details are collected in the
appendices. 

\section{Wilson lines and Reggeization in QCD}
\setcounter{equation}{0}
\label{sec:QCD}

In this section, we review the approach of refs.~\cite{Korchemskaya:1994qp,
Korchemskaya:1996je} for describing the forward limit of $2\rightarrow2$
scattering in QCD in terms of a pair of Wilson lines separated by a transverse
distance. 
Much of this calculation is very similar to the
gravity case considered in the next section, and thus examining the QCD case 
first allows a detailed comparison between gauge and gravity theories. 
Unless otherwise stated, we will consider the scattering of massive particles,
where for convenience we assume a common mass $m$. The Regge
limit we consider is then given by
\begin{equation}
s\gg -t\gg m^2.
\label{limit}
\end{equation}
Note that one has to make a choice here as to how to order the scales $t$ and 
$m^2$, as is inevitable when one introduces a mass scale. It is useful to
have such a mass scale, however, especially when we consider the gravity 
case.\\

The Regge limit corresponds to a high center-of-mass energy, with comparatively
negligible momentum transfer. This corresponds to highly-forward scattering,
such that the incoming particles barely glance off each other. Using the 
momentum labels of figure~\ref{fig:2to2}, the Mandelstam invariants 
are given by eq.~(\ref{mandies}), and momentum conservation can be expressed 
by eq.~(\ref{momcon}). It is clear that in the forward limit the incoming 
particles do not recoil in the transverse direction, and thus can only change 
by a phase due to their interaction. As remarked in the introduction, this 
suggests that one may model the two incoming particles (together with their
outgoing counterparts) by Wilson lines, which are separated by a 
transverse distance $\vec{z}$. The latter is a two-vector which is orthogonal
to the beam direction, corresponding to the {\it impact parameter} or distance
of closest approach. This setup is shown in figure~\ref{fig:Wilson}. 
In principle we need only specify a single direction for each Wilson line. 
However, it is useful to keep the notion of which part of each Wilson line is
incoming and which outgoing, and thus we keep labels for each particle
as shown in the figure.\\

Let us now consider the quantity\footnote{Note that we use a tilde to 
denote the momentum-space Fourier transform of a position-space amplitude.}
\begin{equation}
\widetilde{M}=\int d^2\vec{z}\,e^{-i\vec{z}\cdot\vec{q}}\left\langle 0\right|
\Phi(p_1,0 )\,\Phi(p_2,z)\left|0\right\rangle,
\label{Mdef}
\end{equation} 
where we define the Wilson line operator
\begin{equation}
\Phi(p,z)={\cal P}\exp\left[ig_s\, p^\mu \int_{-\infty}^\infty ds \, A_\mu(sp+z)\right].
\label{phidef}
\end{equation}
The argument of $\Phi$ 
describes the contour of the Wilson line, in 
terms of a momentum $p$ and a constant off-set $z$. The exponent contains
the non-Abelian gauge field $A_\mu$, where the ${\cal P}$ symbol denotes path
ordering of color generators along the Wilson line contour. We then see that 
eq.~(\ref{Mdef}) involves a vacuum expectation value of two Wilson lines along 
directions $p_1$ and $p_2$, separated by the 4-vector $z$.  
As discussed above, this separation will only have non-zero transverse 
components, such that $z^2=-\vec{z}^2$. Were this separation to be absent, the
vacuum expectation value in eq.~(\ref{Mdef}) would correspond exactly to the
Regge limit of the soft function. As is well-known, 
this soft function is exactly zero in dimensional regularization, as it involves
cancellations between UV and IR poles. The former are associated with shrinking
gluon emissions towards the cusp formed by the Wilson lines at the origin.
The presence of the separation vector $\vec{z}$ thus means that the UV poles 
are absent (i.e. there is then no cusp). In other words, $\vec{z}$ acts as a 
UV regulator (see ref.~\cite{Korchemskaya:1994qp} for a prolonged discussion of this
point). \\

Equation~(\ref{Mdef}) constitutes a two-dimensional Fourier transform of the 
Wilson line expectation value, from position to momentum space. The 
two-momentum $\vec{q}$ is conjugate to the impact parameter $\vec{z}$, and 
in fact satisfies $\vec{q}^2=-t$ in the center-of-mass frame. This is because 
in the extreme forward limit, the 4-momentum transfer 
\begin{equation}
q=p_1-p_3
\label{qdef}
\end{equation}
(which will be conjugate to the 4-separation $z$) has zero light-cone 
components
\begin{equation}
q^{\pm}=\frac{1}{\sqrt{2}}(q^0\pm q^3),
\label{qpmdef}
\end{equation}
so that $q=(0,\vec{q},0)$. Our task is now to calculate the quantity of 
eq.~(\ref{Mdef}), and show that it indeed contains known properties of the 
eikonal scattering amplitude.\\

\begin{figure}
\begin{center}
\scalebox{0.7}{\includegraphics{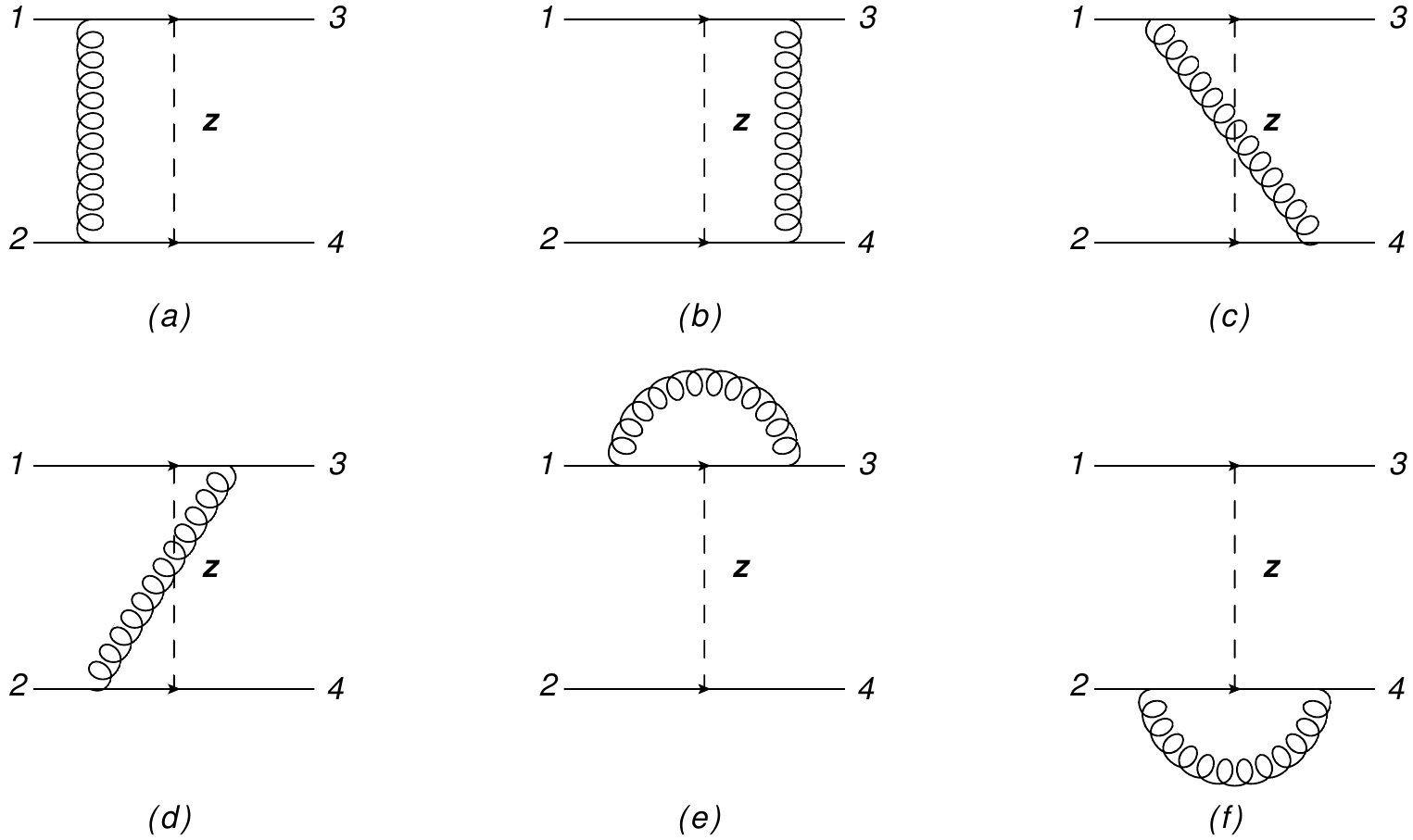}}
\caption{One-loop diagrams entering the calculation of the Wilson line vacuum
expectation value of eq.~(\ref{Mdef}).}
\label{fig:diags}
\end{center}
\end{figure}
The full set of 
one-loop 
diagrams to be calculated is shown in figure~\ref{fig:diags}.
In what follows, we will use the Catani-Seymour notation ${\mathbf T}_i$ to 
denote a color generator on leg $i$~\cite{Catani:1996jh,Catani:1996vz}. 
Using the position space 
gluon propagator (see e.g. ref.~\cite{Korchemsky:1992xv})
\begin{equation}
D_{\mu\nu}(x-y)=-\eta_{\mu\nu}\frac{\Gamma(d/2-1)  }{4\pi^{d/2}}\left[-(x-y)^2\right]^{1-d/2},
\label{prop}
\end{equation}
in $d=4-2\epsilon$ dimensions, diagram (a) gives
\begin{align}
{M}^{(1)}_a&=\frac{g_s^2\Gamma(1-\epsilon) \mu^{2\epsilon}}{4\pi^{2-\epsilon}}{\mathbf T}_1\cdot{\mathbf T}_2
(p_1\cdot p_2)
\int_{-\infty}^0 ds\int_{-\infty}^0 dt\left[
-(sp_1-tp_2)^2+\vec{z}^2\right]^{\epsilon-1}\notag\\
&=\frac{g_s^2\Gamma(1-\epsilon) \mu^{2\epsilon}}{4\pi^{2-\epsilon}}{\mathbf T}_1\cdot{\mathbf T}_2
\cosh\gamma_{12}
\int_{0}^\infty ds
\int_{0}^\infty dt\left[-s^2-t^2+2st\cosh\gamma_{12}+
\vec{z}^2\right]^{\epsilon-1},
\label{Ma1}
\end{align}
where $p_i^2 = m^2$,
the cusp angle $\gamma_{ij}$ is defined via
\begin{equation}
\cosh\gamma_{ij}=\frac{p_i\cdot p_j}{m^2}
\label{gamdef}
\end{equation}
and in the second line of eq.~(\ref{Ma1}) we redefined $s\rightarrow -s/m$, 
$t\rightarrow -t/m$. We also introduced the dimensional regularization scale
$\mu$. 
Next, one may set $s\rightarrow \sqrt{\vec{z}^2}s$, 
$t\rightarrow \sqrt{\vec{z}^2}t$, followed by $t\rightarrow st$, so that
eq.~(\ref{Ma1}) becomes
\begin{align}
{M}^{(1)}_a&=\frac{g_s^2\Gamma(1-\epsilon)}{4\pi^{2-\epsilon}}
{\mathbf T}_1\cdot{\mathbf T}_2(\mu^2\vec{z}^2)^{\epsilon}
\cosh\gamma_{12}\int_{0}^\infty ds\int_{0}^\infty dt \,s\,[s^2(-1-t^2+2t\cosh\gamma_{12})+1]^{\epsilon-1}\notag\\
&=\frac{g_s^2\Gamma(1-\epsilon)}{4\pi^{2-\epsilon}}
{\mathbf T}_1\cdot{\mathbf T}_2(\mu^2\vec{z}^2)^{\epsilon}
\cosh\gamma_{12}\int_{0}^\infty dt\left[\frac{\left[s^2(-1-t^2+2t\cosh\gamma_{12})+1\right]^\epsilon}{2\epsilon(-1-t^2+2t\cosh\gamma_{12})}
\right]^\infty_0.
\label{Ma2}
\end{align}
We see that this result is well-defined for $\epsilon<0$. 
This is to be expected, given that the transverse separation $\vec{z}$
acts as a UV regulator, and $\epsilon$ acts as an IR regulator, 
leaving
\begin{equation}
\label{diagram-a}
{M}^{(1)}_a=\frac{g_s^2\Gamma(1-\epsilon)}{4\pi^{2-\epsilon}}
{\mathbf T}_1\cdot{\mathbf T}_2(\mu^2\vec{z}^2)^{\epsilon}
\frac{1}{2\epsilon}
\cosh\gamma_{12}
\int_0^\infty\frac{dt}{1+t^2-2t\cosh\gamma_{12}}  \,.
\end{equation}
Completing the square in the denominator and substituting
$t = u \sinh \gamma_{12} + \cosh \gamma_{12}$, 
one obtains
\begin{equation}
{M}^{(1)}_a=\frac{g_s^2\Gamma(1-\epsilon)}{4\pi^{2-\epsilon}}
{\mathbf T}_1\cdot{\mathbf T}_2(\mu^2\vec{z}^2)^{\epsilon}
\frac{1}{2\epsilon}
\coth\gamma_{12}
\int_{-\coth\gamma_{12}}^\infty\frac{du}{u^2-1} \,.
\end{equation}
Carefully implementing the $i \varepsilon$ prescription in the propagator, one
evaluates the integral to obtain
\begin{equation}
\int_{-\coth\gamma_{12}}^\infty\frac{du}{u^2-1}
= i\pi-\gamma_{12},
\label{Ma4}
\end{equation}
yielding
\begin{equation}
{M}^{(1)}_a=\frac{g_s^2\Gamma(1-\epsilon)}{4\pi^{2-\epsilon}}
{\mathbf T}_1\cdot{\mathbf T}_2(\mu^2\vec{z}^2)^{\epsilon}
\frac{1}{2\epsilon}(i\pi-\gamma_{12})\coth\gamma_{12}.
\label{Ma5}
\end{equation}
The calculation of diagram (b) in figure~\ref{fig:diags} is very similar, and yields
\begin{equation}
{M}^{(1)}_b=\frac{g_s^2\Gamma(1-\epsilon)}{4\pi^{2-\epsilon}}
{\mathbf T}_3\cdot{\mathbf T}_4(\mu^2\vec{z}^2)^{\epsilon}
\frac{1}{2\epsilon}(i\pi-\gamma_{34})\coth\gamma_{34}.
\label{Mb1}
\end{equation}
Diagrams (c) and (d) are different, because they involve the exchange of a 
gluon between an incoming and outgoing leg, rather than between a pair of
both ingoing (or both outgoing) legs. It is relatively straightforward to 
trace the effect of this in the above calculation;
the effect is to switch the sign of the lower limit of the $u$ integral in 
eq.~(\ref{Ma4}), which then evaluates to $\gamma_{ij}$. 
Thus
\begin{align}
{M}^{(1)}_c&=\frac{g_s^2\Gamma(1-\epsilon)}{4\pi^{2-\epsilon}}
{\mathbf T}_1\cdot{\mathbf T}_4(\mu^2\vec{z}^2)^{\epsilon}
\frac{1}{2\epsilon}\gamma_{14}\coth\gamma_{14};\label{Mc1}\\
{M}^{(1)}_d&=\frac{g_s^2\Gamma(1-\epsilon)}{4\pi^{2-\epsilon}}
{\mathbf T}_2\cdot{\mathbf T}_3(\mu^2\vec{z}^2)^{\epsilon}
\frac{1}{2\epsilon}\gamma_{23}\coth\gamma_{23}\label{Md1}.
\end{align}
Diagram (e) yields
\begin{align}
{M}^{(1)}_e=\frac{g_s^2\Gamma(1-\epsilon)\mu^{2\epsilon}}{4\pi^{2-\epsilon}}{\mathbf T}_1\cdot{\mathbf T}_3
(p_1\cdot p_3)
\int_{-\infty}^0 ds\int_{0}^\infty dt\left[
-(sp_1-tp_3)^2\right]^{\epsilon-1},
\label{Me1}
\end{align} 
which can be obtained from eq.~(\ref{Ma1}) by relabeling of external
momenta and setting the transverse separation $\vec{z}$ to zero. Setting
$s\rightarrow -s/m$ and $t\rightarrow t/m$, this becomes
\begin{align}
{M}^{(1)}_e&=\frac{g_s^2\Gamma(1-\epsilon)\mu^{2\epsilon}}{4\pi^{2-\epsilon}}
{\mathbf T}_1\cdot{\mathbf T}_3\cosh\gamma_{13}\int_{0}^\infty ds\int_{0}^\infty dt\left[-s^2-t^2-2st\cosh\gamma_{13}\right]^{\epsilon-1}\notag\\
&=\frac{g_s^2\Gamma(1-\epsilon)\mu^{2\epsilon}}{4\pi^{2-\epsilon}}
{\mathbf T}_1\cdot{\mathbf T}_3\cosh\gamma_{13}\int_0^\infty ds\,s^{2\epsilon-1}\int_0^\infty dt\,\left[-1-t^2-2t\cosh\gamma_{13}\right]^{\epsilon-1},
\label{Me2}
\end{align} 
where in the second line we have rescaled $t\rightarrow ts$. One sees that the 
$s$ integral contains both a UV and an IR pole. This is to be expected, given 
that there is no transverse separation between particles 1 and 3, 
which acted as a UV regulator in the previous diagrams. One must introduce 
a counterterm for the UV pole, which amounts
to keeping only the IR pole in eq.~(\ref{Me2}). Alternatively, one may simply
introduce a UV cutoff, and here we will use the same cutoff that we have 
already used, returning to this point later. Noting that, after the various 
rescalings we have performed, $s$ has dimensions of length, we may define 
the $s$-integral above via
\begin{equation}
\int_0^\infty ds\,s^{2\epsilon-1}\rightarrow\int_{\sqrt{\vec{z}^2}}^\infty ds\,s^{2\epsilon-1}=-\frac{(\vec{z}^2)^\epsilon}{2\epsilon}.
\label{sint}
\end{equation}
Then eq.~(\ref{Me2}) becomes 
\begin{align}
{M}^{(1)}_e&=\,-\,\frac{g_s^2\Gamma(1-\epsilon)}{4\pi^{2-\epsilon}}
{\mathbf T}_1\cdot{\mathbf T}_3(\mu^2\vec{z}^2)^\epsilon\frac{1}{2\epsilon}
\cosh\gamma_{13}
\int_0^\infty dt\left[-1-t^2-2t\cosh\gamma_{13}\right]^{\epsilon-1}.
\label{Me3}
\end{align}
The remaining integral over $t$ is finite as $\epsilon \to 0$,
in which case it is evaluated similarly to eq.~(\ref{diagram-a}) above 
to give\footnote{Here and in subsequent equations, we keep an overall 
$\epsilon$-dependent factor, which contributes finite terms that
are removed upon renormalizing the Wilson line correlator in the $\msbar$
scheme.}
\begin{equation}
{M}^{(1)}_e=\frac{g_s^2\Gamma(1-\epsilon)}{4\pi^{2-\epsilon}}
{\mathbf T}_1\cdot{\mathbf T}_3(\mu^2\vec{z}^2)^\epsilon\frac{1}{2\epsilon}\gamma_{13}\coth\gamma_{13}
+{\cal O}(\epsilon^0).
\label{Me4}
\end{equation}
Likewise, diagram (f) gives
\begin{equation}
{M}^{(1)}_f=\frac{g_s^2\Gamma(1-\epsilon)}{4\pi^{2-\epsilon}}
{\mathbf T}_2\cdot{\mathbf T}_4(\mu^2\vec{z}^2)^\epsilon\frac{1}{2\epsilon}\gamma_{24}\coth\gamma_{24}
+{\cal O}(\epsilon^0).
\label{Mf1}
\end{equation}
Let us now combine all diagrams and take the Regge limit (\ref{limit}). 
In this limit, we have
\begin{equation}
\gamma_{ij}=\cosh^{-1}\left(\frac{p_i\cdot p_j}{m^2}\right)
\mathrel{\mathop{\longrightarrow}\limits_{p_i\cdot p_j \gg m^2}}
\log\left(\frac{2p_i\cdot p_j}{m^2}\right),
\label{arccosh2}
\end{equation}
and thus, also approximating $(p_i+p_j)^2\simeq 2p_i\cdot p_j$,
\begin{equation}
\gamma_{12},\gamma_{34}\rightarrow\log\left(\frac{s}{m^2}\right),\quad \gamma_{14},\gamma_{23}\rightarrow\log\left(-\frac{u}{m^2}\right),\quad \gamma_{13},\gamma_{24}\rightarrow\log\Big(-\frac{t}{m^2}\Big).
\label{gamlims}
\end{equation}
Furthermore, in the Regge limit one has $s\simeq -u$, so that
\begin{equation}
\gamma_{12},\gamma_{34},\gamma_{14},\gamma_{23}\rightarrow \log\left(\frac{s}{m^2}\right),\quad \gamma_{13},\gamma_{24}\rightarrow\log\Big(-\frac{t}{m^2}\Big).
\label{gamlims2}
\end{equation}
Finally, $\coth(\gamma_{ij})\rightarrow 1$, so that the Regge limit of the
sum of diagrams (a)-(f) gives
\begin{align}
\sum_i{M}^{(1)}_i=\frac{g_s^2\Gamma(1-\epsilon)}{4\pi^{2-\epsilon}}
\frac{(\mu^2\vec{z}^2)^\epsilon}{2\epsilon}\Big\{
&i\pi\left[{\mathbf T}_1\cdot{\mathbf T}_2+{\mathbf T}_3\cdot{\mathbf T}_4\right]\notag\\
+&\log\left(\frac{s}{m^2}\right)
\left[-{\mathbf T}_1\cdot{\mathbf T}_2-{\mathbf T}_3\cdot{\mathbf T}_4
+{\mathbf T}_1\cdot{\mathbf T}_4+{\mathbf T}_2\cdot{\mathbf T}_3\right]\notag\\[3mm]
+&\log\Big(-\frac{t}{m^2}\Big)\left[{\mathbf T}_1\cdot{\mathbf T}_3
+{\mathbf T}_2\cdot{\mathbf T}_4\right]\Big\}
+{\cal O}(\epsilon^0).
\label{Msum}
\end{align}
We may simplify this expression further by introducing the color operators
\begin{equation}
{\mathbf T}^2_s =\left({\mathbf T}_1+{\mathbf T}_2\right)^2, \qquad\qquad
{\mathbf T}^2_t =\left(\mathbf{T}_1-\mathbf{T}_3\right)^2  
\label{ttdef}
\end{equation}
whose eigenstates are pure $s$- and $t$-channel exchanges, and
the corresponding eigenvalue in each case is the quadratic Casimir operator
appropriate to the representation of the exchanged particle. 
Using these, together with color conservation 
${\mathbf T}_1+{\mathbf T}_2={\mathbf T}_3+{\mathbf T}_4$,
we obtain
\begin{align}
&{\mathbf T}_1\cdot{\mathbf T}_2+{\mathbf T}_3\cdot{\mathbf T}_4=
 \frac{1}{2}\left[({\mathbf T}_1+{\mathbf T}_2)^2
+({\mathbf T}_3+{\mathbf T}_4)^2-\sum_{i=1}^4 C_i\right]
={\mathbf T}_s^2-\frac{1}{2}\sum_{i=1}^4 C_i, \\
-&{\mathbf T}_1\cdot{\mathbf T}_2-{\mathbf T}_3\cdot{\mathbf T}_4
+{\mathbf T}_1\cdot{\mathbf T}_4+{\mathbf T}_2\cdot{\mathbf T}_3=
({\mathbf T}_1-{\mathbf T}_3)\cdot({\mathbf T}_4-{\mathbf T}_2) =
{\mathbf T}_t^2, \\
&{\mathbf T}_1\cdot{\mathbf T}_3+{\mathbf T}_2\cdot{\mathbf T}_4=
-\frac{1}{2}\left[({\mathbf T}_1-{\mathbf T}_3)^2
+({\mathbf T}_2-{\mathbf T}_4)^2-\sum_{i=1}^4 C_i\right]
=-{\mathbf T}_t^2+\frac{1}{2}\sum_{i=1}^4 C_i,
\end{align}
where $C_i$ is the quadratic Casimir operator in the representation of 
external particle $i$. 
Using these in eq.~(\ref{Msum}) yields 
\begin{align}
\sum_i{M}^{(1)}_i&=\frac{g_s^2\Gamma(1-\epsilon)}{4\pi^{2-\epsilon}}
\frac{(\mu^2\vec{z}^2)^\epsilon}{2\epsilon}
\left[i\pi{\mathbf T}_s^2 +{\mathbf T}_t^2\log\left(\frac{s}{-t}\right)+
\frac{1}{2}\left(\log\left(-\frac{t}{m^2}\right)-i\pi\right)\sum_{i=1}^4 C_i\right] 
+{\cal O}(\epsilon^0).
\label{Msum2}
\end{align}
Note that we have not included self-energy diagrams associated with any of the
incoming or outgoing particles. These would contribute constant terms which 
will not concern us in what follows. \\

Some further comments are in order regarding the above calculation,
and how it differs from that presented in
refs.~\cite{Korchemskaya:1994qp,Korchemskaya:1996je}. Here, we
separated the incoming and outgoing branch of each Wilson line, and
included diagrams in which a gluon is absorbed and emitted from the
same line, using the same ultraviolet cutoff as for the diagrams in
which a gluon spans both Wilson lines. Had we used a different cutoff,
this would have contributed an additional logarithmic dependence,
beginning only at ${\cal O}(\epsilon^0)$ level. Our motivation for
including the additional diagrams was so as to be able to combine
terms to generate logarithms of $s/(-t)$, as opposed to the
calculation of refs.~\cite{Korchemskaya:1994qp,Korchemskaya:1996je},
which instead considers\footnote{
References~\cite{Korchemskaya:1994qp,Korchemskaya:1996je} 
also consider the alternative Regge limit $s,m^2\gg |t|$, 
rather than the choice made in eq.~(\ref{limit}). 
The diagrams which we include here do not contribute 
logarithms of $s/m^2$ in that paper, so can be neglected.}  
logarithms of $s/m^2$.  The choice made here allows us
to more easily make contact with the case of massless external
particles studied in refs.~\cite{Bret:2011xm,DelDuca:2011ae}, as the
mass dependence has canceled in the color non-diagonal terms. As
$m\rightarrow0$, an additional (collinear) singularity appears in
eq.~(\ref{Msum2}), here appearing as a logarithm of the mass. Were one
to use dimensional regularization 
to regulate both soft and collinear singularities,
eq.~(\ref{Msum2}) would
have a double pole in $\epsilon$ in the massless limit. Because the
above calculation includes soft information only, it also misses hard
collinear contributions, which appear in the full amplitude as (hard)
jet functions divided by {\it eikonal
  jets}~\cite{Mueller:1979ih,Collins:1980ih,Sen:1981sd,Magnea:1990zb}. Such
contributions are irrelevant to the discussion of the Regge
trajectory~\cite{Bret:2011xm,DelDuca:2011ae}, and will not bother us
in gravity, where collinear singularities are
absent~\cite{Weinberg:1965nx,Akhoury:2011kq}.  Apart from the
different collinear regulator, and the lack of hard collinear terms,
eq.~(\ref{Msum2}) agrees with the result found in
refs.~\cite{Bret:2011xm,DelDuca:2011ae} by taking the Regge limit of
the QCD dipole formula at one loop. \\

Note that only the first two 
contributions in the square bracket of eq.~(\ref{Msum2}) have a non-trivial 
color structure when acting on the color structure of the hard interaction. 
The final term is color-diagonal, involving only quadratic Casimir operators.
Let us interpret the various contributions in more detail. 
We know that the soft function exponentiates.
Thus, we may exponentiate eq.~(\ref{Msum2}) to obtain
\begin{align}
\exp\left\{\frac{g_s^2\Gamma(1-\epsilon)}{4\pi^{2-\epsilon}}
\frac{(\mu^2\vec{z}^2)^\epsilon}{2\epsilon}
\left[i\pi{\mathbf T}_s^2 +{\mathbf T}_t^2\log\left(\frac{s}{-t}\right)+
\frac{1}{2}\left(\log\left(-\frac{t}{m^2}\right)-i\pi\right)\sum_{i=1}^4 C_i\right]
\right\}\,.
\label{exp1loop}
\end{align}

In the Regge limit, the term involving $\log(s/-t)$ dominates, and the above
combination reduces to
\begin{equation}
\left(\frac{s}{-t}\right)^{K{\mathbf T}_t^2},\qquad\qquad K=\frac{g_s^2\Gamma(1-\epsilon)}{4\pi^{2-\epsilon}}\frac{(\mu^2\vec{z}^2)^\epsilon}{2\epsilon}.
\label{Reggeop}
\end{equation}
For a Born interaction dominated by a $t$-channel exchange in the
Regge limit (which is usually the case), this operator acts to Reggeize the
exchanged particle. That is, it leads to an amplitude with the behavior
\begin{equation}
\left(\frac{s}{-t}\right)^{J+KC_R},
\end{equation}
where $C_R$ is the quadratic Casimir associated with the exchanged particle,
in representation $R$ of the gauge group, and $J$ is the spin of the particle
(which leads to an appropriate power of $(s/-t)$ in the Born amplitude).\\

However, there are cases in which $C_R=0$. An example is electron scattering in
QED, in which the Born amplitude is dominated by $t$-channel exchange of the
photon, which has zero squared charge. Then the $i\pi$ terms in 
eq.~(\ref{exp1loop}) give
\begin{align}
& \exp\left[\frac{i\alpha}{\epsilon} (\mu^2\vec{z}^2)^\epsilon \right],
\qquad\qquad 
\alpha=\frac{e^2\Gamma(1-\epsilon)}{4\pi^{1-\epsilon}}\coth{\gamma_{12}},
\label{exp1loop2}
\end{align}
where $e$ is the electron charge, and we have restored the full
dependence on the cusp angle $\gamma_{12}=\gamma_{34}$.  Since
\begin{align}
\coth{\gamma_{12}}&=\coth\left[\cosh^{-1}\left(\frac{p_1\cdot p_2}{m^2}\right)\right]
=\frac{s-2m^2}{\sqrt{s(s-4m^2)}},
\label{gam12}
\end{align}
one has 
\begin{equation}
\alpha=\frac{e^2}{4\pi}\frac{s-2m^2}{\sqrt{s(s-4m^2)}} + {\cal O} (\epsilon).
\label{alphaval}
\end{equation}
Equation~(\ref{exp1loop2}) constitutes the QED equivalent of the gravitational {\it eikonal phase} 
discussed in ref.~\cite{Kabat:1992tb}. Expanding in $\epsilon$ gives
\begin{align}
\exp\left[\frac{i\alpha}{\epsilon} (\mu^2\vec{z}^2)^\epsilon \right] &=
\exp\left[i\frac{\alpha}{\epsilon}+i\alpha\log(\mu^2\vec{z}^2)+{\cal O}(\epsilon)\right]
=(\mu^2\vec{z}^2)^{i\alpha}e^{i\alpha/\epsilon}.
\label{exp1loop3}
\end{align}
One can then carry out the Fourier transform of eq.~(\ref{Mdef}) to obtain
(at this order)
\begin{align}
\widetilde{M}=\int d^2\vec{z}e^{-i\vec{q}\cdot \vec{z}}(\mu^2\vec{z}^2)^{i\alpha}e^{i\alpha/\epsilon}
=\frac{4\pi i \alpha }{t} e^{i\alpha/\epsilon}\left(\frac{-t}{4\mu^2}\right)^{-i\alpha}
\frac{\Gamma(1+i\alpha)}{\Gamma(1-i\alpha)}
\label{fourier2}
\end{align}
where we have taken a Hankel transform of order zero, and recalled that $t = -\vec{q}^2$.
This has poles in the plane of the Mandelstam invariant $s$, stemming from 
the $\Gamma$ function in the numerator i.e. when
\begin{equation}
i\alpha=-N,\quad N=1,2,\ldots
\label{poles}
\end{equation}
Then eq.~(\ref{alphaval}) implies that the physical poles of the scattering amplitude are at
\begin{equation}
s=2m^2\left[1-\left(1+\frac{e^4}{16\pi^2 N^2}\right)^{-1/2}\right].
\label{poles2}
\end{equation}
Given that poles in $s$ of a scattering amplitude represent bound states,
eq.~(\ref{poles2}) represents the spectrum of $s$-channel states 
produced in electron scattering (i.e. positronium). Indeed, the above 
calculation reproduces eq. (17) of ref.~\cite{Brezin:1970zr}. \\

In this section, we have introduced the Wilson line formalism 
of ref.~\cite{Korchemskaya:1994qp} for examining the Regge limit in QCD and QED. 
In particular, we have seen two effects emerge:
\begin{itemize}
\item If the Born interaction is dominated by a $t$-channel exchange in the
Regge limit, then this particle Reggeizes at leading log  order, with a trajectory which
depends on the quadratic Casimir in the appropriate representation of the 
gauge group.
\item There is a pure phase term, the {\it eikonal phase}, which is associated
with the formation of $s$-channel bound states.
\end{itemize}
As is well-known, the first of these contributions arises at the Feynman 
diagram level from vertical ladder graphs, and the second arises from 
horizontal ladder graphs. Which of these is kinematically leading in the
Regge limit depends in the present case on the squared charge of the particle
being exchanged in the $t$-channel. If this is non-zero, the Reggeization
term dominates. If however, the squared charge is zero (as in the case of the
photon), then the eikonal phase is the dominant effect. \\

Things get more complicated beyond leading logarithmic order in $(s/-t)$. 
One must include higher order contributions to the soft function, as well as
include the possibility of cross-talk between the eikonal phase and 
Reggeization terms. This becomes especially cumbersome in QCD, due to the fact
that the color operators associated with the eikonal phase and Reggeization
terms do not commute. This has already been noted 
in refs.~\cite{Bret:2011xm,DelDuca:2011ae}, where it was
identified with a lack of simple Regge pole behavior at NNLL order. \\

Having seen how things work in QED and QCD, we examine the case of gravity in
the following section.

\section{Wilson line approach for gravity}
\setcounter{equation}{0}
\label{sec:gravity}

In the previous section, we have reviewed the Wilson line approach for 
examining the Regge limit 
of gauge theory scattering amplitudes
in some detail. 
The case of gravitational scattering can be obtained quite straightforwardly from
the above results. Note that we here discuss explicitly the case of 
Einstein-Hilbert gravity. As we will see, this will also have features in 
common with supersymmetric extensions.\\

Let us first recall the form\footnote{The 
factor of two error in eq.~(3.2) of 
ref.~\cite{Naculich:2011ry} has been corrected in v3.}
of the gravitational Wilson line operator \cite{Naculich:2011ry,White:2011yy}
(see also ref.~\cite{Brandhuber:2008tf})
\begin{equation}
\Phi_g(p,z)=\exp\left[i\frac{\kappa}{2}p^\mu p^\nu\int_{-\infty}^{\infty} ds\, h_{\mu\nu}(sp+z)\right],
\label{phigdef}
\end{equation}
where $\kappa=\sqrt{32\pi G_N}$ in terms of Newton's constant $G_N$.
We will use the de Donder gauge graviton propagator\footnote{Note that this
differs by a factor of $-2$ from that used in ref.\cite{Miller:2012an}, 
which can be traced to our use of $\kappa=\sqrt{32\pi G_N}$ 
and metric $(+,-,-,-)$ in the present paper, 
rather than $\kappa=\sqrt{16\pi G_N}$ and metric $(-,+,+,+)$ 
in ref.~\cite{Miller:2012an}.}
\begin{equation}
D_{\mu\nu,\alpha\beta}(x-y)={P_{\mu\nu,\alpha\beta}}\frac{\Gamma(d/2-1)}{4\pi^{d/2}}\left[-(x-y)^2\right]^{1-d/2}, \quad 
P_{\mu\nu,\alpha\beta}=
\frac{1}{2} \left( \eta_{\mu\alpha}\eta_{\nu\beta}+\eta_{\nu\alpha}\eta_{\mu\beta}-\frac{2}{d-2}\eta_{\mu\nu}\eta_{\alpha\beta}\right).
\label{propgrav}
\end{equation}
By analogy with the gauge theory case, we now wish to calculate the amplitude
\begin{equation}
\widetilde{M}_g=\int d^2\vec{z}\,e^{-i\vec{z}\cdot\vec{q}}\left\langle 0\right|\Phi_g(p_1,0)\,\Phi_g(p_2,z)\left|0\right\rangle,
\label{Mgdef}
\end{equation} 
i.e. a pair of gravitational Wilson lines separated by a transverse distance
$\vec{z}$. The diagrams will be the same as those of figure~\ref{fig:diags}. 
Given that the denominator structure of the propagator~(\ref{propgrav}) is 
the same as that of~(\ref{prop}), we do not have to recalculate any of the 
kinematic integrals. All that changes in each diagram is the overall prefactor
of $p_i\cdot p_j$, obtained by contracting two eikonal Feynman rules with 
the gluon propagator. In the gravity case this will be replaced by
\begin{equation}
p_i^\mu\,p_i^\nu\,P_{\mu\nu,\alpha\beta}\,p_j^\alpha\,p_j^\beta=
(p_i\cdot p_j)^2-\frac{1}{d-2}m^4.
\label{eikruleg}
\end{equation}

The result for diagram (a) is then
\begin{equation}
{M}^{(1)}_{g,a}=~-~\left(\frac{\kappa}{2}\right)^2\frac{\Gamma(1-\epsilon)}{4\pi^{2-\epsilon}}
(\mu^2 \vec{z}^2)^{\epsilon}\left[p_1\cdot p_2-\frac{m^4}{2(1-\epsilon)}\frac{1}{p_1\cdot p_2}\right]\frac{1}{2\epsilon}(i\pi-\gamma_{12})\coth\gamma_{12}.\label{Mga1}
\end{equation}
In the Regge limit, neglecting terms of ${\cal O}(m^2/s,m^2/t)$, 
this may be obtained from eq.~(\ref{Ma5}) by replacing 
\begin{equation}
g_s\rightarrow\frac{\kappa}{2},
\qquad\qquad
{\mathbf T}_i \rightarrow p_i
\end{equation}
and switching the overall sign.
The other diagrams are similar, so the 
sum of gravitational diagrams in the Regge limit may be obtained by
making the replacements (as $m\rightarrow 0$)
\begin{equation}
g_s\rightarrow\frac{\kappa}{2}, \qquad
{\mathbf T}^2_s \to s,  \qquad {\mathbf T}^2_t \to t, \qquad C_i \to 0 
\label{simplereplace}
\end{equation} 
and switching the overall sign in eq.~(\ref{Msum2}), yielding
\begin{equation}
\sum_i{M}^{(1)}_{g,i}
= ~-~\left(\frac{\kappa}{2}\right)^2\frac{\Gamma(1-\epsilon)}{4\pi^{2-\epsilon}}
\frac{(\mu^2\vec{z}^2)^\epsilon}{2\epsilon}\left[ i \pi s + t\log\left(\frac{s}{-t}\right)\right] 
+{\cal O}(\epsilon^0)
\label{Msumgrav}
\end{equation}
where we have dropped the ${\cal O}(m^2)$ terms which vanish in the Regge limit.
Note that 
the logarithmic dependence on the mass has completely canceled
in the sum over diagrams due to the absence of collinear divergences in 
gravity \cite{Weinberg:1965nx}.
Thus one would expect the same result for the scattering 
of strictly massless particles.
Furthermore, one-loop exactness tells us that there are no perturbative 
corrections to eq.~(\ref{Msumgrav}).\\

We see that two terms occur in the soft function in the Regge limit: 
an $i \pi s$ eikonal phase term, and a  $ t \log(s/-t)$ term, 
which will Reggeize the graviton.
Expanding eq.~(\ref{Msumgrav}) in $\epsilon$ and exponentiating the result gives
\begin{equation}
e^{-i\pi sK_g/\epsilon}\left(\frac{s}{-t}\right)^{-K_gt/\epsilon}
\left(\mu^2\vec{z}^2\right)^{-K_g[i \pi s + t\log(s/-t)]},
\qquad\qquad 
K_g=\left(\frac{\kappa}{2}\right)^2\frac{\Gamma(1-\epsilon)}{8\pi^{2-\epsilon}}.
\label{expg}
\end{equation}
When acting on the Born interaction (which is ${\cal O}(s^2)$), 
the power-like term in $(s/-t)$ corresponds to Reggeization of the graviton with
a trajectory
\begin{equation}
\alpha_g(t)=2-\frac{tK_g}{\epsilon}.
\label{alphagdef}
\end{equation}
We see that the one-loop perturbative Regge trajectory in gravity is 
infrared singular (up to scale logarithms), as is known to be the case in QCD
and QED. However, this trajectory is linear in the Mandelstam invariant $t$,
reproducing the results of refs.~\cite{Lipatov:1982it,Lipatov:1982bf,Lipatov:1982vv,
Schnitzer:2007kh,Schnitzer:2007rn}. In the present approach, however, the
comparison with QCD appears in a particularly elegant fashion. As discussed 
in ref.~\cite{Bret:2011xm,DelDuca:2011ae}, we expect Reggeization to occur at 
leading logarithmic order if the tree level interaction is dominated by the
$t$-channel exchange of a particle with a non-vanishing squared charge. 
The Regge trajectory is then infrared singular at one loop, and 
contains the quadratic Casimir associated with the exchanged particle. Here
exactly the same mechanism occurs for the graviton, and the relevant 
gravitational quadratic Casimir is the squared four-momentum, which in this 
case is simply the Mandelstam invariant $t$. \\

The eikonal phase term, which in QCD involved a quadratic Casimir operator for $s$-channel 
exchanges, now contains the Mandelstam invariant $s$. This in turn implies
that Reggeization of the graviton is kinematically suppressed with respect to
the eikonal phase in the strict Regge limit of $s/(-t)\rightarrow\infty$. 
In Feynman diagram terms: horizontal ladders and crossed ladders (which build 
up the eikonal phase as discussed in ref.~\cite{Kabat:1992tb}) win out over 
vertical ladders (which build up the Reggeized graviton). Nevertheless, both 
effects are present and clearly show up in the Wilson line calculation.  \\

As has already been commented above, in hindsight we could have
obtained the gravity result of eq.~(\ref{Msumgrav}) from the QCD case
of eq.~(\ref{Msum2}) without detailed calculation, by the simple
replacements of eq.~(\ref{simplereplace}).  The final replacement
corresponds to the setting to zero of quadratic Casimir operators
associated with the external legs, here a consequence of having
considered massless particles in the gravity case (indeed, as
discussed in ref.~\cite{Oxburgh:2012zr}, this is one way of
appreciating the cancellation of collinear divergences in gravity).
The replacements (\ref{simplereplace}) are consistent, at least in
general, with what one would expect from the double copy procedure of
refs.~\cite{Bern:2008qj,Bern:2010ue,Bern:2010yg}.  In addition to the
coupling constant replacement, color operators are replaced by their
momentum counterparts which, in Feynman diagram language, is
equivalent to the replacement of color factors by kinematic
numerators. The double copy was considered in more detail in this
context in ref.~\cite{Saotome:2012vy}, which also discussed the
relationship between shock waves in both gauge and gravity
theories. The latter point can also be understood in the language of
Wilson lines, as we briefly describe in appendix~\ref{app:shock}.\\

As in the QED case, one may carry out the Fourier transform over the impact
parameter. After substituting eq.~(\ref{expg}) into eq.~(\ref{Mgdef}), one 
obtains (at this order)
\begin{align}
\widetilde{M}_g&=\frac{-4\pi K_ge^{-i\pi sK_g/\epsilon}}{t}\left(\frac{s}{-t}\right)^{-K_gt/\epsilon}
\left[i \pi s + t\log\left(\frac{s}{-t}\right)\right]
\left(\frac{-t}{4\mu^2}\right)^{K_g[i \pi s + t\log(s/-t)]}\notag\\
&\quad\times\frac{\Gamma[1-K_g(i\pi s+t\log(s/-t))]}{\Gamma[1+K_g(i\pi s+t\log(s/-t))]}.
\label{Mgres}
\end{align}
The ratio of Euler gamma functions no longer constitutes a pure phase. Also,
it now gives rise to cuts in the $s$ plane, rather than poles. By standard
Regge theory arguments, the high-energy behavior of an amplitude $A(s,t)$ is 
related to its analytically continued partial wave coefficients $F(t,j)$,
where the angular momentum $j$ has become a complex variable, by 
(see e.g. ref.~\cite{SUSSP})
\begin{equation}
F(t,j)=\int_1^\infty ds\, s^{-j-1} A(s,t).
\label{Frel}
\end{equation}
Thus, cuts in the $s$-plane give rise to Regge cuts in the complex angular 
momentum plane. Note that such cuts will only appear if both
the Reggeization and eikonal phase term are kept, thus they are due to 
a cross-talk between these two contributions. This is consistent with the
results of refs.~\cite{Bret:2011xm,DelDuca:2011ae}, which demonstrated a breakdown
of Regge pole behavior at three loop order in QCD, associated with the
presence of both a Reggeization and an eikonal phase term. This was assumed
to herald the arrival of Regge cut contributions at this order in perturbation
theory. For example, the color factor associated with the non-Regge-pole-like
contribution was non-planar, and consistent with Feynman diagrams which lead 
to cuts~\cite{Mandelstam:1965zz}. Here we see directly that cross-talk between
the eikonal phase and Reggeization terms leads to cut-like behavior. 
It is interesting to remark that gravity theories provide a simpler testing 
ground for such ideas, lying somewhere between Abelian and non-Abelian gauge 
theories in terms of complication: although multi-graviton vertices are 
present (unlike an Abelian gauge theory), there is no non-commuting color 
structure. There may well be other problems in QCD whose conceptual structure 
is simplified by examining a gravitational analogue.\\

If we neglect the Reggeization term, and restore full mass dependence in the 
eikonal phase term, eq.~(\ref{Mgres}) becomes
\begin{align}
\widetilde{M}_g
&=~-~ \frac{4\pi i G(s)    }{t} e^{-iG(s)/\epsilon}
\left(\frac{-t}{4\mu^2}\right)^{iG(s)}
\frac{\Gamma[1-iG(s)]}{\Gamma[1+iG(s)]},
\label{Mgres2}
\end{align}
where\footnote{Note that the quantity $G(s)$ is referred to as $\alpha(s)$
in ref.~\cite{Kabat:1992tb}.}
\begin{equation}
G(s)={G_N}\left(\frac{s^2-4m^2s+2m^4}{\sqrt{s(s-4m^2)}}\right)
\label{tildeKdef}
\end{equation}
which essentially agrees with the eikonal amplitude 
in refs.~\cite{'tHooft:1987rb,Kabat:1992tb}.
The Euler gamma function then gives rise to poles in the amplitude, 
corresponding to the spectrum of bound states discussed in section IV 
of ref.~\cite{Kabat:1992tb}.\\

In this section, we have seen that both a Reggeization term and an eikonal
phase term are present in gravity. However, the Regge trajectory of the 
graviton is linear in $t$, and hence the eikonal phase dominates in the 
strict Regge limit. Cross-talk between the eikonal phase and Reggeization 
terms is associated with Regge cut behavior.\\

The above analysis was carried out in Einstein-Hilbert (non-supersymmetric)
gravity. However, it also applies to the four-graviton amplitude in 
supergravity, if one dresses only the tree-level hard interaction with the 
eikonal calculation discussed here.
This is because the leading infrared singularity at each order in perturbation
theory arises from the Born amplitude dressed only by graviton emissions 
between the external legs (the highest spin objects in the theory). There
are no corrections to the gravitational soft function, as dictated by 
one-loop exactness~\cite{Weinberg:1965nx,Naculich:2011ry,White:2011yy,
Akhoury:2011kq}. However, subleading IR singularities (and infrared finite 
parts) will arise in the amplitude from higher order contributions to the 
hard interaction, which are sensitive to the additional matter content, and
hence the degree of supersymmetry. \\

The only information that we have used about the hard interaction in the above
calculation is that the Wilson lines are separated by a transverse distance.
This means that we have no control over finite parts of the amplitude. 
One would think this is irrelevant to the issue of graviton Reggeization at 
one-loop order, as the perturbative Regge trajectory is purely infrared 
singular at this order. However, the finite terms do lead to complications, as
we discuss in the following section.

\section{Infrared-finite contributions in supergravity}
\setcounter{equation}{0}
\label{sec:finite}

In the previous sections, we have reviewed the Regge limit of QCD from a Wilson
line point of view, and applied this same reasoning to gravity. Use of this
common language showed a number of similarities between the two theories:
namely the presence of both a Reggeization and eikonal phase term, and an 
infrared singular Regge trajectory at one loop that contained the relevant
quadratic Casimir operator. These facts, by themselves, lead to the fact that
the eikonal phase is kinematically dominant in gravity, and subdominant in QCD.
In this section, we discuss another important difference between the QCD and
gravity cases: in the latter, Reggeization is interrupted even at one loop by
the presence of double log terms of the same order in $x\equiv-t/s$ in
the infrared finite part of the amplitude. Let us begin by considering the
one-loop amplitude. 

\subsection{One-loop results}
\label{sec:oneloop}
We here consider ${\cal N}=M$ supergravity, where $4\leq M\leq 8$.
One-loop results were obtained in 
refs.~\cite{Green:1982sw,Dunbar:1994bn,Dunbar:1995ed,Bern:2011rj}.
Following ref.~\cite{BoucherVeronneau:2011qv}, we write
the one-loop four-graviton amplitude as
\begin{equation}
{\cal M}_4^{(1),\,{\cal N}=M}=
\left( \frac{\kappa}{8\pi} \right)^2
\left(\frac{4\pi\,e^{-\gamma_E}\,\mu^2}{|s|}\right)^\epsilon
\left\{\frac{2}{\epsilon}\big[s\log(-s)+t\log(-t)+u\log(-u)\big]+F_4^{(1),\, {\cal N}=M}\right\}
{\cal M}_4^{\rm tree},
\label{M41def}
\end{equation}
where $\kappa=\sqrt{32\pi G_N}$,
$\gamma_E$ is Euler's constant,
and 
$F_4^{(1),\, {\cal N}=M}$ is an IR-finite contribution 
dependent on the degree of supersymmetry.  \\

The infrared-singular part of  eq.~(\ref{M41def}), 
as remarked in the previous section, 
is universal at one-loop order.
In the physical region $s>0$;  $t,u<0$,
it is given by 
\begin{equation}
{\cal M}_4^{(1),\,{\cal N}=M} \Big|_{\rm IR-divergent} = 
\left( \frac{\kappa}{8\pi} \right)^2
\frac{2}{\epsilon}
\big[ s\log s-i\pi s+t\log(-t)+u\log(-u)\big]
{\cal M}_4^{\rm tree}
\end{equation}
where $\log(-s)=\log|s|-i\pi$.
Setting $u=-s-t$, and expanding about the Regge limit $s \gg -t$,
one obtains
\begin{equation}
{\cal M}_4^{(1),\,{\cal N}=M} \Big|_{\rm IR-divergent}
=
~-~ \frac{\kappa^2}{32\pi^2 \epsilon} 
\left[i\pi s + t\log\left(\frac{s}{-t}\right) +t + {\cal O} \left(\frac{t^2}{s}\right) \right]
{\cal M}_4^{\rm tree}
\end{equation}
As expected, 
this agrees with the result (\ref{Msumgrav}) obtained from the Wilson line 
calculation (up to the non-logarithmic ${\cal O}(t)$ term neglected in the latter).  \\

Next we consider the IR-finite part of the amplitude in the Regge limit.
The Regge limit corresponds to $x \to 0$ with $s$ fixed, where 
\begin{equation}
x \equiv  \frac{-t}{s}.
\label{xdef}
\end{equation}
The various remainder terms $F_4^{(1)}$ for 
different supergravity theories are collected 
in the appendix of ref.~\cite{BoucherVeronneau:2011qv}. 
We substitute these into eq.~(\ref{M41def}), set $u = -t-s$,
and keep the first two terms in the expansion about $x=0$ to obtain 
(for $s>0$, $t<0$)
\begin{align}
{\cal M}_4^{(1),\,{\cal N}=8}&=\left(\frac{\kappa}{8\pi}\right)^2\left(\frac{4\pi\,e^{-\gamma_E}\,\mu^2}{-t}\right)^\epsilon\bigg\{\frac{s}{\epsilon}\left[-2i\pi+2x(L+1)\right]+sx\left[-2L^2+2i\pi L\right]\bigg\}{\cal M}_4^{\rm tree};\notag\\
{\cal M}_4^{(1),\,{\cal N}=6}&=\left(\frac{\kappa}{8\pi}\right)^2\left(\frac{4\pi\,e^{-\gamma_E}\,\mu^2}{-t}\right)^\epsilon\bigg\{\frac{s}{\epsilon}\left[-2i\pi+2x(L+1)\right]+sx\left[-L^2+2i\pi L+\pi^2\right]\bigg\}{\cal M}_4^{\rm tree};\notag\\
{\cal M}_4^{(1),\,{\cal N}=5}&=\left(\frac{\kappa}{8\pi}\right)^2\left(\frac{4\pi\,e^{-\gamma_E}\,\mu^2}{-t}\right)^\epsilon\bigg\{\frac{s}{\epsilon}\left[-2i\pi+2x(L+1)\right]+sx\left[-\frac{L^2}{2}+2i\pi L+\frac{3\pi^2}{2}\right]\bigg\}{\cal M}_4^{\rm tree};\notag\\
{\cal M}_4^{(1),\,{\cal N}=4}&=\left(\frac{\kappa}{8\pi}\right)^2\left(\frac{4\pi\,e^{-\gamma_E}\,\mu^2}{-t}\right)^\epsilon\bigg\{\frac{s}{\epsilon}\left[-2i\pi+2x(L+1)\right]+sx\left[2\pi i L - L +2\pi^2 + 1\right]\bigg\}{\cal M}_4^{\rm tree},
\label{M4res}
\end{align}
where for convenience we define $L=\log(s/-t)$. 
These results may be compactly summarized as
\begin{align}
{\cal M}_4^{(1),\,{\cal N}=M}
&=\left(\frac{\kappa}{8\pi}\right)^2\left(\frac{4\pi\,e^{-\gamma_E}\,\mu^2}{-t}\right)^\epsilon
\bigg\{\frac{s}{\epsilon}\left[-2i\pi+2xL + 2x\right]\notag\\
&\quad+sx\left[\left(\frac{4-M}{2}\right)L^2+\left(\frac{8-M}{2}\right)\pi^2+2i\pi L+\delta_{M4}(1-L)\right]+ {\cal O}(s x^2) + {\cal O}(\epsilon)
\bigg\}{\cal M}_4^{\rm tree}, 
\label{M4res2}
\end{align}
which makes clear the dependence on the degree of supersymmetry $M$. \\

The first two terms in the infrared-singular part of eq.~(\ref{M4res2}), as discussed at
length in the previous section, correspond to the eikonal phase and 
Reggeization of the graviton respectively, where the latter is kinematically 
suppressed (${\cal O}(x)$ in the present notation). However, the first term of
the infrared-finite part contains the double log contribution (ignoring prefactors)
\begin{equation}
sx\left(\frac{4-M}{2}\right)L^2 =  \left(\frac{M-4}{2}\right)t\log^2\left(\frac{s}{-t}\right)
\end{equation}
as observed in ref.~\cite{Bartels:2012ra}.
This does not correspond to Reggeization of the graviton which, as we have 
already seen, is purely infrared singular at this order and can involve only
a single log. Nevertheless, the double logarithmic contribution is of the 
same order (linear in $x$) as the Reggeization term, and in fact 
superleading (logarithmically in $x$) with respect to the Regge logs. The fact
that the coefficient of the double logarithmic contribution is sensitive to the
additional matter content of the theory (via the degree of supersymmetry $M$) 
tells us that one is not sensitive to this contribution in the Wilson line
approach, which picks up only graviton-related contributions at one loop
(this also explains why the double log is in the infrared finite part).\\

\begin{figure}
\begin{center}
\scalebox{1.0}{\includegraphics{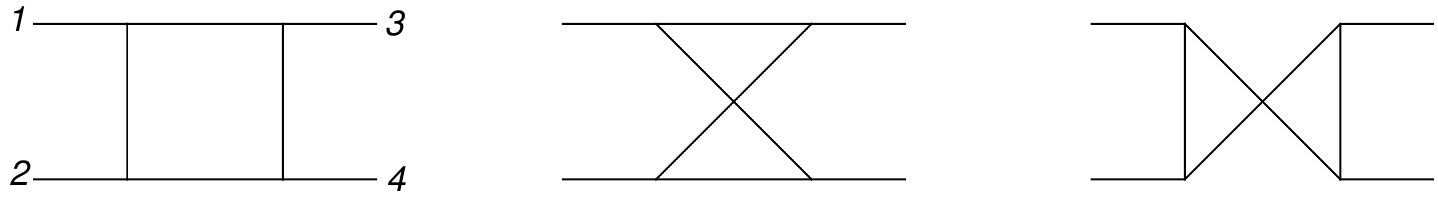}}
\caption{Diagrams contributing to the one-loop four-graviton amplitude in ${\cal N}=8$ supergravity.}
\label{fig:boxes}
\end{center}
\end{figure}

Another way to see that the double logs at one loop are not associated with 
Reggeization is to examine their origin in terms of the Feynman diagrams
contributing to the amplitude. Taking the example of ${\cal N}=8$ supergravity,
the one-loop amplitude may be written as~\cite{Bern:1998ug}
\begin{equation}
{\cal M}_4^{(1),\, {\cal N}=8}=-i\left(\frac{\kappa}{2}\right)^2stu\left[{\cal I}_4^{(1)}(s,t)+{\cal I}_4^{(1)}(t,u)+{\cal I}_4^{(1)}(s,u)\right]
{\cal M}_4^{\rm tree},
\label{M4I}
\end{equation}
where
\begin{equation}
{\cal I}_4^{(1)}(s,t)=\mu^{4-d}\int\frac{d^dk}{(2\pi)^d}\frac{1}{k^2\,(k-p_2)^2\,(k-p_2-p_1)^2\,(k+p_4)^2}
\label{I4def}
\end{equation}
is the first scalar box integral shown in figure~\ref{fig:boxes}. The terms
${\cal I}_4^{(1)}(t,u)$ and ${\cal I}_4^{(1)}(s,u)$ 
then correspond to the
second and third box diagrams in the figure. 
The result for the integral may be written~\cite{Bern:2005iz}
\begin{equation}
{\cal I}_4^{(1)}(s,t)=\frac{ie^{-\epsilon\gamma_E}(4\pi)^{\epsilon-2}}{st}
\left\{\frac{4}{\epsilon^2}-\frac{2}{\epsilon}\log\left(\frac{-s}{\mu^2}\right)-\frac{2}{\epsilon}\log\left(\frac{-t}{\mu^2}\right)+2\log\left(\frac{-s}{\mu^2}\right)\log\left(\frac{-t}{\mu^2}\right)-\frac{4\pi^2}{3}+\mathcal{O}(\epsilon)\right\}
\label{I4res}
\end{equation}
in $d=4-2\epsilon$ dimensions in the region $s$, $t<0$. From this result, we see that
the double logarithmic contribution comes from ${\cal I}_4^{(1)}(s,u)$, 
corresponding to the third diagram in figure~\ref{fig:boxes}. This is neither
a ladder nor a crossed ladder, and thus is responsible neither for the eikonal
phase, nor for the Reggeization of the graviton. \\

The question then arises how to interpret the additional double 
logarithmic contributions, and whether or not they exponentiate.
This has been discussed in ref.~\cite{Bartels:2012ra}, which argues for two sources 
of double logarithms. The first is from ladder contributions, including 
infrared-finite effects of the graviton Regge 
trajectory~\cite{Lipatov:1982it,Lipatov:1982bf,Lipatov:1982vv}. The second is
that backward-scattering contributions are important at this order in $t/(-s)$,
an observation corroborated by the fact that such contributions arise from the
dressed $u$-channel diagram, the third in fig.~\ref{fig:boxes}. 
The authors of ref.~\cite{Bartels:2012ra} write an evolution equation for the leading 
partial wave contributing to the amplitude in the limit in which double 
logarithms are important, whose solution is argued to resum these 
contributions. Note, however, that such contributions become more and more
kinematically suppressed at higher orders in perturbation theory, giving rise
to terms
\begin{equation}
\left[\kappa^2 t\log^2\left(\frac{s}{-t}\right)\right]^n
\sim \kappa^{2n}s^{n}x^nL^{2n},
\end{equation}
which are ${\cal O}(x^n)$. 

\subsection{Two-loop results}

In the previous section, we have seen that the interpretation of Reggeization
of the graviton is interrupted at one-loop order by the presence of double
logarithmic contributions in the infrared finite part of the amplitude, which
are the same order in $t/s$ as the Reggeization terms. This motivates an 
examination of the four-graviton amplitude at two loops, with a view to 
seeing which structures exponentiate, and which do not.\\

Following ref.~\cite{BoucherVeronneau:2011qv}, we write the two-loop amplitude as
\begin{equation}
\frac{{\cal M}_4^{(2),\,{\cal N}=M}(\epsilon)}{{\cal M}_4^{\rm tree}}=\frac{1}{2}\left[\frac{{\cal M}_4^{(1),\,{\cal N}=M}(\epsilon)}{{\cal M}_4^{\rm tree}}\right]^2+\left(\frac{\kappa}{8\pi}\right)^4F_4^{(2),\,{\cal N}=M}+{\cal O}(\epsilon),
\label{F4def}
\end{equation}
where the IR-finite remainder function $F_4^{(2),\,{\cal N}=M}$ 
corresponds to the part of the two-loop result that is not generated by 
exponentiation of the one-loop result\footnote{As the notation in 
eq.~(\ref{F4def}) suggests, in constructing the remainder one must be mindful 
of terms generated due to the cross-talk between ${\cal O}(\epsilon)$ and 
${\cal O}(\epsilon^{-1})$ terms when squaring the one-loop result.}.
The remainder function for ${\cal N}=8$ supergravity
was computed in  refs.~\cite{Naculich:2008ew,Brandhuber:2008tf}
using the results of  refs.~\cite{Bern:1998ug,Smirnov:1999gc,Tausk:1999vh},
and for ${\cal N} = M < 8$ supergravity in ref.\cite{BoucherVeronneau:2011qv}.
Again defining $x=-t/s$ and expanding about the Regge limit $x\rightarrow 0$
(keeping terms up to linear in $x$),
one finds that the behavior of each remainder function is 
\begin{align}
F_4^{(2),\,{\cal N}=8}=s^2 x\bigg\{&-2\pi^2\log^2 x-4\pi^2\log x +\pi^4+4\pi^2
\notag\\
&
+i\pi\left[\frac{4}{3}\log^3 x + 4\log^2x -\left(8+\frac{8\pi^2}{3}\right)\log x + 16\zeta_3+\frac{8\pi^2}{3}+8\right]\bigg\}+\cdots\label{m48}\\
F_4^{(2),\,{\cal N}=6}=s^2 x\bigg\{&-2\pi^2\log^2 x-4\pi^2\log x +\frac{59\pi^4}{90}+4\pi^2
\notag\\
&+i\pi\left[\frac{2}{3}\log^3 x + 4\log^2x
-\left(8+\frac{6\pi^2}{3}\right)\log x + 4\zeta_3+\frac{16\pi^2}{3}+8\right]\bigg\}+\cdots\label{m46}\\
F_4^{(2),\,{\cal N}=5}=s^2 x\bigg\{&-2\pi^2\log^2 x-4\pi^2\log x +\frac{2\pi^4}{3}+4\pi^2
\notag\\
&+i\pi\left[\frac{1}{3}\log^3 x + 4\log^2x
-\left(8+\frac{5\pi^2}{3}\right)\log x + 4\zeta_3+\frac{20\pi^2}{3}+8\right]\bigg\}+\cdots\label{m45}\\
F_4^{(2),\,{\cal N}=4}=s^2 x\bigg\{&-2\pi^2\log^2 x-4\pi^2\log x +\frac{13\pi^4}{30}+\frac{22\pi^2}{3}-1
\notag\\
&+i\pi\left[3\log^2x
-\left(14+\frac{4\pi^2}{3}\right)\log x - 4\zeta_3+\frac{71\pi^2}{9}+\frac{32}{3}\right]\bigg\}+\cdots.\label{m44}
\end{align}
Note that the IR-finite remainder functions vanish in the 
strict Regge limit $x\rightarrow0$. 
This is because the amplitude in this limit is dominated by the eikonal phase dressing the 
tree-level result. The eikonal phase contribution exponentiates (at least) up to this 
order, and thus the remainder must vanish in the limit. \\

One may summarize the {\it logarithmic} terms of the remainder function, 
for general $M$, as
\begin{align}
F_4^{(2),\,{\cal N}=M}=s^2 x\bigg\{&-2\pi^2\log^2 x-4\pi^2\log x
\notag\\&+i\pi\left[\left(\frac{M-4}{3}\right)\log^3 x + (4-\delta_{M4})\log^2x
-\left(8+\frac{M\pi^2}{3}+6\delta_{M4}\right)\log x \right]\bigg\}
+ \cdots 
\label{mgen}
\end{align}
where, as in the one-loop amplitude, $M=4$ is a somewhat exceptional
case, due presumably to the decreasing amount of supersymmetry as one counts
down from $M=8$.
In both the one- and two-loop amplitudes, 
a number of terms are independent of $M$, 
and thus are common to all the supergravity
theories considered here. Such terms presumably arise from finite contributions
involving the graviton alone. At two loops such contributions would be 
ultraviolet divergent in pure Einstein-Hilbert gravity. Here, the results are 
made finite by the additional matter content of the various supergravities.\\

In eqs.~(\ref{m48}-\ref{m44}),  
we have not displayed terms of ${\cal O}(s^2 x^2)$, 
but as noted in ref.~\cite{Bartels:2012ra}, 
the remainder function can contain quartic logarithms at this order
\begin{equation}
F_4^{(2),\,{\cal N}=M}
~=~ \cdots  ~+~ 
c_M s^2 x^2 \log^4 \left( \frac{s}{-t} \right) ~+~ \cdots, \qquad\qquad \mbox{where}  \quad
\begin{cases} 
c_8 = -\frac{1}{3} \\ c_6 = 0 \\ c_5 = \frac{1}{24} \\ c_4 = 0
\end{cases} 
\end{equation}
showing that, for $M=5$ and $M=8$, 
the one-loop double logarithmic terms do not formally exponentiate. 
The authors of ref.~\cite{Bartels:2012ra} argue that these terms can be resummed
to all orders. 
In any case, 
eq.~(\ref{mgen}) makes clear that there is a more dominant source
of IR-finite corrections at two-loop order, namely those which are 
${\cal O}(s^2 x)$. \\

These ${\cal O}(s^2 x)$ terms also threaten a simple interpretation of Reggeization 
at this order, as they introduce a dependence on $s/(-t)$ which is 
kinematically enhanced relative to the Reggeization of the graviton at this 
order. It is interesting to ponder whether any of the terms in the above 
remainders can be shown to exponentiate, or be resummable in some other form. 
It is known, for example, that $t/(-s)$ corrections to the eikonal phase 
should come into play in describing black-hole 
formation~\cite{Amati:1992zb,Giddings:2010pp}.
That such features are suppressed in this manner is partly due to the fact that
they are not described by the eikonal approximation, which reproduces the
bound states associated with only the perturbative (Coulomb-like) part of the 
gravitational potential. Black holes should be associated with non-perturbative
dynamics, as discussed in ref.~\cite{Kabat:1992tb}. \\

Because the two-loop remainder functions vanish in the strict Regge limit 
$x\rightarrow 0$ for arbitrary degrees of supersymmetry, 
the four-graviton scattering amplitude is reproduced exactly in this limit
by the exponentiation of the one-loop result at two-loop order. 
One may wonder whether this remains true at higher orders. 
To this end, it is interesting to note that the eikonal result itself
does not satisfy this requirement at three-loop order and beyond. To see this,
note that the ratio of Euler gamma functions in eq.~(\ref{Mgres2}) can be
expanded to give
\begin{equation}
\frac{\Gamma[1-iG]}{\Gamma[1+iG]}
=e^{ 2 i\gamma_E G}
\left[1+\frac{i}{3}\Psi^{(2)}(1)G^3+{\cal O}(G^4)\right],
\label{Gamexp}
\end{equation}
where $\Psi^{(n)}(x)$ is the $n^{\rm th}$ derivative of the digamma function
\begin{equation}
\Psi(x)=\frac{d}{dx}\log\Gamma(x).
\label{psidef}
\end{equation}
Equation~(\ref{Gamexp}) does not have a purely exponential form, and
shows that one-loop exactness of the Regge limit of the amplitude may
be broken at three-loop level and beyond by infrared-finite
contributions.  This is not a firm conclusion, given that there may be
infrared-finite corrections to the amplitude which are not captured by
the eikonal approximation which leads to eq.~(\ref{Gamexp}). However,
the ratio of gamma functions resums contributions of known physical
origin (the formation of bound states in the $s$-channel), and so
presumably describes genuine behavior to all orders in perturbation
theory.

\section{The Regge limit of multigraviton amplitudes}
\setcounter{equation}{0}
\label{sec:multi}

\begin{figure}
\begin{center}
\scalebox{0.8}{\includegraphics{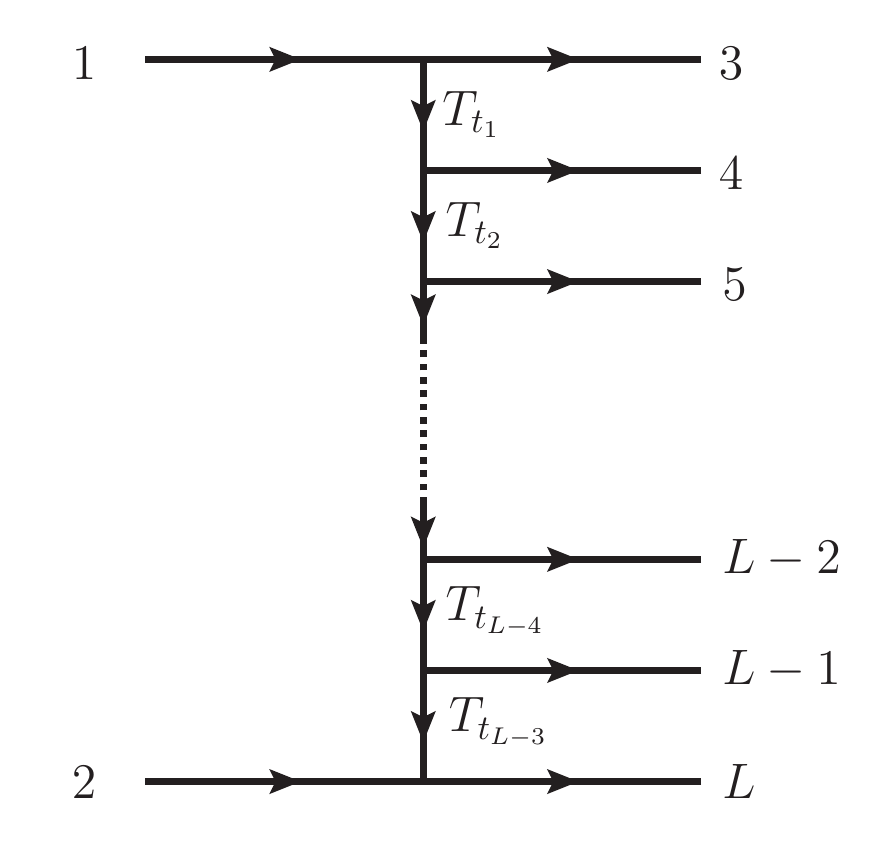}}
\caption{A general $L$-parton scattering process in the MRK limit,
consisting of strongly ordered rapidities in the final state. Here $T_{t_i}$ is
a quadratic Casimir operator associated with a given strut of the ladder.}
\label{MRfig}
\end{center}
\end{figure}

In previous sections, we have considered the four-graviton scattering 
amplitude, consisting of $2\rightarrow 2$ scattering dressed by virtual 
graviton exchanges. The Regge limit has also been widely studied for the case
of general $L$-point scattering, with $L>4$ 
(for a pedagogical review in a QCD context, see ref.~\cite{DelDuca:1995hf}). This
was studied from an infrared point of view 
in refs.~\cite{Bret:2011xm,DelDuca:2011ae}, which confirmed the result that in the
high energy limit, scattering is dominated by multiple $t$-channel exchanges,
as shown in figure~\ref{MRfig}, where each strut of the ladder is dressed by
a Reggeized propagator involving the relevant quadratic Casimir 
for the exchanged object.\\

Given the results of section~\ref{sec:gravity} for the $2\rightarrow 2$
scattering in gravity, it is instructive to examine the high energy limit of
multigraviton scattering using the Wilson line approach. As for the four-point
amplitude, this provides an interesting comparative study with respect to
non-Abelian gauge theory. Furthermore, it is useful to clarify the role of the
eikonal phase in this context.\\

First, let us briefly review the QCD case, differing 
from refs.~\cite{Bret:2011xm,DelDuca:2011ae} in that we use a Wilson line 
calculation, rather than the dipole formula~\cite{Gardi:2009qi,
Becher:2009cu,Becher:2009qa} as a starting point. We will consider massive 
particles, keeping a mass $m$ only when regulating collinear singularities
(as is done in e.g. eq.~(\ref{Msum})). For ease of comparison 
with refs.~\cite{Bret:2011xm,DelDuca:2011ae} (and also for the sake of brevity in
the following formulae), we will reverse the sign of the color
generators associated with the incoming legs i.e. 
${\mathbf T}_{1,2}\rightarrow -{\mathbf T}_{1,2}$. Treating each external leg
of the amplitude as a separate Wilson line, the set of all contributing
one-loop diagrams consists of gluon emissions between pairs of external lines.
From eq.~(\ref{Msum}), one infers that the sum of these diagrams 
gives (including the color adjustment mentioned above)
\begin{align}
{M}_L^{(1)}&=\frac{g_s^2\Gamma(1-\epsilon)}{4\pi^{2-\epsilon}}
\frac{(\mu^2\,\Lambda_{UV}^2)^\epsilon}{2\epsilon}\left\{i\pi\left({\mathbf T}_1\cdot{\mathbf T}_2+\sum_{i=3}^{L-1}\sum_{j>i}{\mathbf T}_i\cdot{\mathbf T}_j\right)
-{\mathbf T}_1\cdot {\mathbf T}_2\log\left(\frac{s}{m^2}\right)\right.\notag\\
&\left.\quad
-\sum_{i=3}^{L-1}\sum_{i>j}{\mathbf T}_i\cdot{\mathbf T}_j\log\left(\frac{s_{ij}}{m^2}\right)-\sum_{i=3}^L\left[{\mathbf T}_1\cdot{\mathbf T}_i\log\left(-\frac{s_{1i}}{m^2}\right)+{\mathbf T}_2\cdot{\mathbf T}_i\log\left(-\frac{s_{2i}}{m^2}\right)\right] \right\},
\label{Msummult}
\end{align}
where
\begin{equation}
s = (p_1+p_2)^2, \quad
s_{ij}=(p_i+p_j)^2,\quad s_{1i}=(p_1-p_i)^2,\quad s_{2i}=(p_2-p_i)^2; \quad i,j>3,
\label{sijdef}
\end{equation}
and $\Lambda_{UV}^2$ is an ultraviolet regulator, which we have chosen to be
the same for all diagrams. We may now use the fact that the high energy limit
of multiparton scattering corresponds to the {\it multi-Regge-kinematic} (MRK)
regime in which the outgoing particles are widely separated in rapidity. One
may then replace the various invariants appearing in eq.~(\ref{Msummult}) 
with (see e.g. ref.~\cite{DelDuca:1995hf})
\begin{align}
s&\simeq |k_{3\perp}|\,|k_{L\perp}|e^{y_3-y_L},\notag\\
-s_{1i}&\simeq |k_{3\perp}|\,|k_{i\perp}|e^{y_3-y_i},\notag\\
-s_{2i}&\simeq |k_{L\perp}|\,|k_{i\perp}|e^{y_L-y_i},\notag\\
s_{ij}&\simeq |k_{i\perp}|\,|k_{j\perp}|e^{y_i-y_j},\quad 3\leq i<j\leq L,
\label{sijlims}
\end{align}
where $y_i$ and $k_{i\perp}$ are the rapidity and transverse momentum of 
parton $i$ respectively. Furthermore, given that the separation between all
pairs of consecutive final state particles is asymptotically approaching 
infinity, this suggests that one may identify the common ultraviolet cutoff
(motivated by the $2\rightarrow 2$ case) with the impact parameter $\vec{z}$
corresponding to the distance of closest approach of the incoming particles. 
In any case, different cutoff choices will not affect the infrared behavior,
only contributing additional logarithms in the infrared finite part of 
eq.~(\ref{Msummult}).\\

Substituting eq.~(\ref{sijlims}) into eq.~(\ref{Msummult}), one may rewrite the
latter as
\begin{align}
{M}_L^{(1)}&=\frac{g_s^2\Gamma(1-\epsilon)}{4\pi^{2-\epsilon}}
\frac{(\mu^2\,\vec{z}^2)^\epsilon}{2\epsilon}\left\{-\sum_{i=1}^{L-1}\sum_{j>i}
|y_i-y_j|{\mathbf T}_i\cdot{\mathbf T}_j+i\pi\left({\mathbf T}_1\cdot{\mathbf T}_2+\sum_{i=3}^{L-1}\sum_{j>i}{\mathbf T}_i\cdot {\mathbf T}_j\right)\right.\notag\\
&\quad\left.+\sum_{i=1}^L C_i\log\left(\frac{|k_{i\perp}|}{m}\right)\right\},
\label{Msummult2}
\end{align}
where $C_i={\mathbf T}_i^2$ is the quadratic Casimir in the representation of
leg $i$, and we made repeated use of the color conservation equation
\begin{equation}
\sum_{i=1}^L {\mathbf T}_i=0.
\label{colcon2}
\end{equation}
Also in eq.~(\ref{Msummult2}), we have introduced the (unphysical) rapidities 
$y_1\equiv y_3$, $y_2\equiv y_L$, in order to simplify the notation.
Introducing the $s$-channel quadratic Casimir
\begin{equation}
{\mathbf T}^2_s=\left({\mathbf T}_1+{\mathbf T}_2\right)^2=\left(\sum_{i=3}^L{\mathbf T}_i\right)^2,
\label{Tsdef2}
\end{equation}
one may also write eq.~(\ref{Msummult2}) as
\begin{align}
{M}_L^{(1)}&=\frac{g_s^2\Gamma(1-\epsilon)}{4\pi^{2-\epsilon}}
\frac{(\mu^2\,\vec{z}^2)^\epsilon}{2\epsilon}\left\{-\sum_{i=1}^{L-1}\sum_{j>i}
|y_i-y_j|{\mathbf T}_i\cdot{\mathbf T}_j+i\pi{\mathbf T}_s^2+\sum_{i=1}^L C_i\left[\log\left(\frac{|k_{i\perp}|}{m}\right)-\frac{i\pi}{2}\right]\right\}.
\label{Msummult3}
\end{align}
One may now use the identity, proven in ref.~\cite{DelDuca:2011ae},
\begin{equation}
\sum_{i=1}^{L-1}\sum_{j>i}|y_i-y_j|{\mathbf T}_i\cdot{\mathbf T}_j=-\sum_{k=3}^{L-1}{\mathbf T}_{t_{k-2}}^2\Delta y_k,
\label{Tid}
\end{equation}
where ${\mathbf T}_{t_i}$ is a quadratic Casimir operator for a given strut of
the $t$-channel ladder in figure~\ref{MRfig}, and $\Delta y_k=y_k-y_{k+1}$ the
associated rapidity difference. Equation~(\ref{Msummult3}) then becomes
\begin{align}
{M}_L^{(1)}&=\frac{g_s^2\Gamma(1-\epsilon)}{4\pi^{2-\epsilon}}
\frac{(\mu^2\,\vec{z}^2)^\epsilon}{2\epsilon}\left\{\sum_{k=3}^{L-1}{\mathbf T}_{t_{k-2}}^2\Delta y_k+i\pi{\mathbf T}_s^2+\sum_{i=1}^L C_i\left[\log\left(\frac{|k_{i\perp}|}{m}\right)-\frac{i\pi}{2}\right]\right\},
\label{Msummult4}
\end{align}
which differs from the result in refs.~\cite{Bret:2011xm,DelDuca:2011ae} owing to
the use of a mass regulator for collinear singularities adopted here. As usual,
one may exponentiate the one-loop soft function to obtain
\begin{align}
\exp\left\{\frac{g_s^2\Gamma(1-\epsilon)}{4\pi^{2-\epsilon}}
\frac{(\mu^2\,\vec{z}^2)^\epsilon}{2\epsilon}\left\{\sum_{k=3}^{L-1}{\mathbf T}_{t_{k-2}}^2\Delta y_k+i\pi{\mathbf T}_s^2+\sum_{i=1}^L C_i\left[\log\left(\frac{|k_{i\perp}|}{m}\right)-\frac{i\pi}{2}\right]\right\}\right\},
\label{Msummult5}
\end{align}
which acts on the hard interaction consisting of $L$-point scattering undressed
by virtual emissions. The leading high energy behavior (corresponding to 
leading logarithms in rapidity) is given by the first term in the exponent 
acting on the hard function, and produces a tower of Reggeized gluon exchanges,
each dressed by the appropriate quadratic Casimir. The analysis is in fact 
more general than this - even if the struts of the ladder have different 
exchanges, each will Reggeize separately given that the $t$-channel operators
for different struts commute with each other~\cite{DelDuca:2011ae}. Note that
an eikonal phase term remains present, weighted as in the $2\rightarrow 2$ 
case by the quadratic Casimir operator for $s$-channel exchanges.
This has the same physical meaning in the present context - it is associated
with the formation of $s$-channel bound states.\\

Having reviewed the QCD case, let us now return to gravity. As may be confirmed
by more detailed calculation, the latter case is easily obtained from the 
former using the replacements of eq.~(\ref{simplereplace}), so that the 
exponentiated gravitational soft function is 
\begin{align}
\exp\left\{\,-\, \left(\frac{\kappa}{2}\right)^2\frac{\Gamma(1-\epsilon)}{4\pi^{2-\epsilon}}
\frac{(\mu^2\,\vec{z}^2)^\epsilon}{2\epsilon}\left[\sum_{k=3}^{L-1}t_{k-2}\Delta y_k+i\pi s\right]\right\}.
\label{Msummultgrav}
\end{align}
Here $t_k$ is the 
squared
momentum transfer flowing in the $k^{\rm th}$ strut of the 
ladder, as labeled in figure~\ref{MRfig}. Here one sees a similar story to 
the QCD case, namely the presence of both a Reggeization and an eikonal phase
term. The former is now itself a series of terms, each of which Reggeizes the
graviton in a given strut of the ladder. However, as in the $2\rightarrow 2$
case, the Reggeization term in gravity involves the squared momentum transfer
by virtue of its being the appropriate quadratic Casimir, and thus is 
kinematically subleading in the strict Regge limit of $s/|t|\rightarrow\infty$.
Thus, multigraviton scattering for any number of gravitons is dominated by the
eikonal phase term, hinting at the production of bound states in the 
$s$-channel.\\

It would be interesting to consider the impact of infrared finite corrections
on this result. By analogy with the four point amplitude, one would expect
additional logarithms to appear in the finite part, which disrupt the 
interpretation of graviton Reggeization. It is worth noting here also that
the Regge limit of multiparticle scattering has been widely investigated in 
the context of the BDS conjecture~\cite{Bern:2005iz}, an all-order ansatz for 
the form of planar amplitudes in ${\cal N}=4$ super-Yang-Mills theory. This
conjecture is known to break down for six-point amplitudes at two loops,
as first shown by considering the Regge limit in an unphysical 
region~\cite{Bartels:2008ce}. One might expect similar structures to occur
in (super)-gravity theories, using double 
copy~\cite{Bern:2008qj,Bern:2010ue,Bern:2010yg} considerations. 

\section{Conclusions}
\setcounter{equation}{0}
\label{sec:conclude}

In this paper, we have considered the Regge limit of gravity from a Wilson line
point of view, adopting an approach first used for Abelian and non-Abelian
gauge theories~\cite{Korchemskaya:1994qp,Korchemskaya:1996je}. Our motivation
was to provide a common way of looking at Reggeization in different theories,
and to clarify the role of graviton Reggeization as presented in the 
literature.\\

The Wilson line approach reveals the presence of both an eikonal phase and a
Reggeization term in the soft function at one-loop, where the former is 
associated with the formation of $s$-channel bound states due to the 
perturbative part of the potential. In QCD, the Reggeization term dominates 
at leading logarithmic order, leading to automatic Reggeization of arbitrary
$t$-channel exchanges, as discussed in refs.~\cite{Bret:2011xm,DelDuca:2011ae}, where
the Regge trajectory is purely infrared singular at one-loop order, and 
involves the quadratic Casimir operator associated with a given $t$-channel 
exchange. Beyond this logarithmic order, cross-talk occurs between the two 
contributions, leading to a breakdown of simple Regge pole behavior. The 
situation is further complicated in QCD, even for the purely infrared singular
parts of the amplitude, by the presence of corrections to the exponent of the
soft function.\\

Our gravity calculation used the Wilson line operators 
of refs.~\cite{Naculich:2011ry,White:2011yy}, and confirmed the presence of both
an eikonal phase and Reggeization term at one-loop in gravity. Here the soft
function is one-loop exact, receiving no perturbative corrections in the 
exponent~\cite{Weinberg:1965nx,Naculich:2011ry,White:2011yy,Akhoury:2011kq}.
The eikonal phase and Regge trajectory, as expected from the QCD calculation,
contain quadratic Casimir operators associated with $s$- and $t$-channel 
exchanges. In the gravity case, these are the Mandelstam 
invariants
$s$ and $t$
themselves, and thus one finds a particularly elegant explanation for the
fact that the gravitational Regge trajectory is linear in $t$, and thus
kinematically-subleading 
in the Regge limit 
with respect to the eikonal phase 
$i \pi s$. 
We saw that 
cross-talk between the two contributions leads to Regge cut behavior, 
clarifying the QCD discussion of refs.~\cite{Bret:2011xm,DelDuca:2011ae}. 
We 
also examined 
Reggeization in multigraviton scattering, finding again that
graviton Reggeization is sub-dominant with respect to the eikonal phase.\\

The story of Reggeization in gravity is further complicated, 
even at one-loop order, 
by the presence of IR-finite $\log^2 s$ contributions.
Such double-log terms arise in explicit one-loop calculations 
in $\cN=5$, $\cN=6$, and $\cN=8$ supergravity \cite{Bartels:2012ra}.
Although these terms are kinematically-suppressed with respect
to the eikonal phase term, they are of the same order as the
Reggeization term and therefore mix with the Reggeization of the graviton. \\

Using known results for the two-loop amplitude in 
in $\cN=4$, $\cN=5$, $\cN=6$, and $\cN=8$ supergravity,
we computed the Regge limit of the two-loop contribution 
to the logarithm of the amplitude, 
which measures the failure of the one-loop result to exponentiate.
These correction terms are of ${\cal O}(st)$ and therefore
kinematically-subleading with respect to the ${\cal O}(s^2)$
exponentiation of the eikonal phase 
(and hence vanish in the strict Regge limit).  
They are, however, kinematically-superleading with respect
to the  ${\cal O}(t^2)$ exponentiation of the 
one-loop Reggeization term and also with respect
to the ${\cal O}(t^2 \log^4 s)$ terms computed
in ref.~\cite{Bartels:2012ra}.\\

Although the strict Regge limit of the two-loop amplitude 
was shown to be one-loop-exact for $\cN \ge 4$ supergravity,
it remains an open question whether this continues to hold
at three loops and beyond, i.e.
whether the strict Regge limit of the $L$-loop result 
is given by the exponential of the one-loop eikonal phase.
The resolution of this question awaits the evaluation of the
contributing non-planar integrals. \\

It is fair to say that the higher-loop contributions to 
the Regge limit of gravity are still not fully understood. 
Investigation of these contributions in more detail may shed 
light on a number of unresolved issues
in quantum gravity, including issues of black hole formation and 
unitarity (see e.g. refs.~\cite{Giddings:2010pp,Giddings:2009gj,Giddings:2011xs}
and references therein).

\section*{Acknowledgments}
We are grateful to Jochen Bartels and Einan Gardi for useful
discussions, and to David Miller for comments on the manuscript.  CDW
thanks Brandeis University, where part of this work was carried out,
for warm hospitality. Both he and SM are supported by the UK Science
and Technology Facilities Council (STFC).  The research of SGN is
supported in part by the NSF under grant no. PHY10-67961.  The
research of HJS is supported in part by the DOE under grant
DE-FG02-92ER40706.

\appendix

\section{Wilson lines and shockwaves}
\setcounter{equation}{0}
\label{app:shock}

The authors of ref.~\cite{Saotome:2012vy} examined the Regge limit of quantum gravity
from the point of view of the double copy procedure of refs.~\cite{Bern:2008qj,
Bern:2010ue,Bern:2010yg}, which posits that gravitational scattering amplitudes
can be obtained from gauge theory counterparts, by replacements of kinematic
numerators by color factors (together with relevant coupling constants). That
paper also pointed out that shockwave solutions - namely gauge field 
configurations corresponding to a single massless particle - can also be 
related by the double copy. In this appendix, we briefly state how shockwaves
are connected to the Wilson line language.\\

Consider first a QED Wilson line operator
\begin{equation}
\exp\left[ie\int dx^\mu A_\mu(x)\right],
\label{QEDWilson}
\end{equation}
with the contour chosen to be the classical straight-line trajectory of a hard 
emitting particle
\begin{equation}
x^\mu=u^\mu \tau,
\label{trajectory}
\end{equation}
where $\tau$ is a parameter along the contour (with units of length), and 
$u^\mu=p^\mu/E$ the 4-velocity of a massless particle with energy $E$. We may
rewrite eq.~(\ref{QEDWilson}) in terms of a current density sourcing the 
gauge field, by introducing a three-dimensional delta function as follows:
\begin{equation}
\exp\left[ie\int dx^\mu A_\mu(x)\right]=
\exp\left[ieu^\mu\int d^4x \, \delta^{(3)}(\vec{x})A_\mu(x)\right]\equiv
\exp\left[-i\int d^4x j^\mu(x)A_\mu(x)\right],
\label{QEDWilson2}
\end{equation}
where
\begin{equation}
\delta^{(3)}(\vec{x})=\delta(z-t)\delta(x)\delta(y),
\label{delta}
\end{equation}
and without loss of generality we have taken the Wilson line to be in the $+z$
direction. We then see that the current due to the Wilson line operator is
\begin{equation}
j^\mu=-eu^\mu\delta(z-t)\delta(x)\delta(y).
\label{jQED}
\end{equation}
As pointed out in ref.~\cite{Saotome:2012vy}, this is precisely the source that 
gives rise to a QED shockwave, upon solving the field equations for 
the gauge field $A_\mu(x)$.\\

A similar argument may be made for gravity, and one starts by rewriting the
Wilson line operator of eq.~(\ref{phigdef}) as\footnote{Strictly speaking,
in gravity there should be factors of $\sqrt{-g}$ in the volume measure, where
$g$ is the determinant of the metric tensor. However, these can be ignored in
eq.~(\ref{gravwilson}) due to the fact that we are only expanding to first 
order in the graviton field.}
\begin{align}
\exp\left[i\frac{\kappa}{2}p^\mu\int dx^\mu h_{\mu\nu}(x)\right]&=
\exp\left[i\frac{\kappa}{2}Eu^\mu\,u^\nu\int d^4 x 
\delta^{(3)}(\vec{x})
h_{\mu\nu}(x)\right]
=\exp\left[-i\int d^4 x j^{\mu\nu}(x)h_{\mu\nu}(x)\right].
\label{gravwilson}
\end{align}
We recognize the source current in this case as
\begin{align}
j^{\mu\nu}(x)&=-\frac{\kappa}{2}Eu^\mu\, u^\nu\delta(z-t)\delta(x)\delta(y)
\equiv -\frac{\kappa}{2}T^{\mu\nu},
\label{jgrav}
\end{align}
where we have introduced the conventional energy-momentum tensor in the 
last term.
This can be recognized as the energy-momentum tensor for a massless 
particle quoted in ref.~\cite{Saotome:2012vy}, so that solution of the field 
equations for $h_{\mu\nu}(x)$ gives the Aichelberg-Sexl (shockwave) metric.

\vfil\break
\providecommand{\href}[2]{#2}\begingroup\raggedright\endgroup

\end{document}